\documentclass[11pt]{article}
\usepackage[margin=1in]{geometry}
\usepackage{amsmath,amssymb,amsthm}
\usepackage{graphicx}
\usepackage{booktabs}
\usepackage{natbib}
\usepackage{hyperref}
\usepackage{xcolor}
\usepackage{enumitem}
\usepackage{caption}
\usepackage{subcaption}
\usepackage{float}

\hypersetup{colorlinks=true,linkcolor=blue!60!black,citecolor=blue!60!black,urlcolor=blue!60!black}

\newtheorem{definition}{Definition}

\title{\textbf{ The Rise and Fall of $G$ in AGI}\\[6pt]}

\author{David C.\ Krakauer\\
\small Santa Fe Institute, 1399 Hyde Park Road, Santa Fe, NM. 87501. USA} 

\begin{document}
\maketitle

\begin{abstract}
In the psychological literature the term `general intelligence' describes correlations between abilities and not simply the number of abilities. This paper connects Spearman's $g$-factor from psychometrics, measuring a positive manifold, to the implicit ``$G$-factor'' in claims about artificial general intelligence (AGI) performance on temporally structured benchmarks. By treating LLM benchmark batteries as cognitive test batteries and model releases as subjects, principal component analysis is applied to a models $\times$ benchmarks $\times$ time matrix spanning 39 models (2019--2025) and 14 benchmarks. Preliminary results confirm a strong positive manifold in which all 28 pairwise correlations positive across 8 benchmarks. By analyzing the spectrum of the benchmark correlation  through time, PC1 explains 90\% of variance on a 5-benchmark core battery ($n=19$)) reducing to 77\% by 2024. On a four benchmark battery,  PC1 is found to peak at 92\% of the variance between 2023--2024 and reduce to 64\% with the arrival of reasoning-specialized models in 2024. This is coincident with a rotation in the G-factor as models outsource `reasoning' to tools.  The analysis of partial correlation matrices through time provides evidence for the evolution of specialization beneath the positive manifold of general intelligence (AI-hedgehog) encompassing diverse high dimensional problem solving systems (AI-foxes). In strictly psychometric terms, AI models exhibit general intelligence  suppressing specialized intelligences. LLMs invert the ideal of substituting complicated models with parsimonious mechanisms, a `Ptolemaic Succession' of theories,  with architectures of increasing hierarchical complication and capability. 
\end{abstract} 


\section{Analytics of Generality}

\subsection{The Positive Manifold }

In 1904, Spearman observed that scores on diverse cognitive tests were positively correlated and proposed a single latent variable $g$, ``general intelligence''---to account for this \emph{positive manifold} \citep{spearman1904}. The observation was robust: across batteries of tests, populations, and cultures, the intercorrelation matrix of cognitive tasks is overwhelmingly positive, and a single first principal component typically accounts for 40--60\% of variance \citep{jensen1998,carroll1993}.

The case for $g$ rests on convergent evidence. The statistical regularity is well established: the positive manifold has been replicated across diverse test batteries and populations \citep{johnson2008}. Carroll's survey of factor-analytic studies, building on Cattel's typology \citep{cattell1963theory}  established a hierarchy---$g$ at the top, broad abilities (fluid reasoning $G_f$, crystallized knowledge $G_c$, and others) in the middle and narrow abilities at the base \citep{carroll1993}. And $g$ is a strong predictor of job performance, educational attainment, and health outcomes across occupations and contexts \citep{gottfredson1997}. Evidence also exists for $g$ in mice \citep{matzel2003individual}.   Different test batteries, constructed independently and measuring ostensibly different abilities, yield $g$-factors that correlate at $r > 0.95$ \citep{johnson2004}. Jensen's characterization of $g$ as a ``distillate'' of cognitive performance extracted by factor analysis from a limited set of tasks captures both its narrowness and  its consistency \citep{jensen2002}.

Challenges to the theoretical significance of $g$ are numerous and include the fact that it tends to collapse informative factors or abilities \citep{thurstone1938} \citep{gardner1983}. It places excessive emphasis on analytical forms of reasoning over creative and practical reasoning \citep{sternberg1985, gottfredson2003}, and arises easily from overlaps in sampling processes \citep{thomson1916, bartholomew2009}. Moreoever $g$ is  expected to grow as a natural consequence of overlapping processes \citep{kovacs2016} and developmental mutualisms \citep{vandermaas2006}. A recent extension to these synergistic frameworks is that of Savi et al.\ in which intelligence is conceptualized as an evolving graph of densely connected facts and procedures \citep{savi2019}.

\subsection{Psychometric $g$ vs Psychological $g$}

The debate centers around a distinction that is relevant to the analysis of LLM benchmarks. \emph{Psychometric $g$} (the first principal component of a battery of cognitive tests) is a statistical regularity. \emph{Psychological $g$}, understood as a causal entity or a single cognitive capacity that explains why tests correlate, is a theoretical possibility. Spearman's notion of ``mental energy,'' Jensen's characterization of $g$ as processing speed or neural efficiency, and the parieto-frontal integration theory's identification of $g$ with a specific brain network all represent claims about psychological $g$ \citep{jung2007}. Thomson's sampling theory, mutualism, and process overlap theory  accept a limited form of psychometric $g$ while denying psychological $g$. As Savi et al.\ put it, psychometric $g$ is an index that summarizes a system without causing it \citep{savi2019}. And the correlation versus causation debate has been extended to general intelligence across many species \citep{burkart2017evolution}.

\subsection{From $g$ to $G$: The AGI Debate}

The AI community has conducted a parallel debate about ``general intelligence'' in tests on artificial systems, largely without reference to the test of human intelligence literature. This is not a criticism but a surprising observation based on the fact, as described above, that the psychometric literature does have a definition of `general intelligence'.  The central claim that large language models exhibit a form of general intelligence was made explicitly by Bubeck et al.\ in ``Sparks of Artificial General Intelligence,'' which showed that GPT-4 could solve novel tasks spanning mathematics, coding, medicine, and law. Hendrycks et al.\ \citep{hendrycks2025agi} proposed a quantitative definition of AGI grounded in the Cattell--Horn--Carroll (CHC) theory defining AGI as an AI that matches or exceeds the cognitive versatility and proficiency of a well-educated adult. LLMs are typically evaluated on batteries of benchmarks---MMLU \citep{hendrycks2021mmlu}, GSM8K \citep{cobbe2021gsm8k}, HumanEval \citep{chen2021humaneval}, GPQA \citep{rein2024gpqa}, MATH \citep{hendrycks2021math}---and performance across these batteries is taken as evidence for or against ``general'' intelligence. Responses to the ``Sparks'' claim have divided along lines that parallel the psychometric debate about $g$. Chollet argued that benchmark performance measures \emph{skill}, not \emph{intelligence}, and that skill can be purchased with sufficient training data without implying any general reasoning capacity \citep{chollet2019}. Morris et al.\ have proposed a ``Levels of AGI'' framework that attempted to operationalize the concept by distinguishing depth (performance on specific tasks) from breadth (generality across tasks), and argued that current systems occupy a position of ``Competent'' narrow AI rather than any level of genuine AGI \citep{morris2023}.

\subsection{Temporal Psychometrics of Transformers}

This paper pursues an opportunity provided by LLMs: the positive manifold can be observed as models evolve in response to benchmarks.  The LLM setting recapitulates the psychometric structure as models are evaluated on benchmark batteries and performance is positively correlated across tasks.  The ``subjects'' (models) are released in a known temporal order with documented architectural differences, so the eigen-structure can be tracked across algorithmic epochs. Benchmarks saturate and are replaced on a timescale of months rather than decades, making the moving-battery problem tractable. And the distinction between statistical and mechanistic $G$ can be mapped onto the Spearman-vs-Thomson debate: if the positive manifold in LLM benchmarks reflects a shared inferential mechanism (the transformer architecture, the language-modeling objective), it is closer to a mechanistic $G$; if it reflects the shared training corpus or the trivial temporal trend of later-models-being-better, it is closer to a statistical $G$.

 The claim that a model is approaching AGI should be a straightforward claim about the time-dependent structure of its benchmark correlation matrix. In time a single dominant eigenvalue, the Principal Component, should come to account for cross-task performance. This paper makes the analogy explicit,  and by dividing benchmarks into distinct model epochs,  asks whether it is substantive. It is argued that a more pragmatic approach might be better off adopting the language of the `Dimension of Intelligence', which is quantitatively justifiable, and allows for different forms of intelligence, animal or artificial, to occupy different subspaces of competence.
 
In prior work Ilic and Gignac \citep{ilic2024evidence} found strong evidence for a positive manifold in LLM performance, with a dominant leading $g$-factor accounting for around 66 percent of test variance. 
Ilic and Gignac analyze 12 benchmarks organized into four broad-ability categories using confirmatory factor analysis and conclude that performance on benchmarks justifies describing models in terms of achievement rather than intelligence.  They propose that Artificial General Achievement (AGA) is a good fit to Cattell's crystallized intelligence and find little evidence for fluid intelligence.  This paper builds on the idea that there is value in exploring how psychological metrics might describe AI performance; in this case using principal components to explore correlations across benchmarks with an emphasis on the changing nature of the positive manifold through time.

\section{Formal Framework}

\subsection{The Data Structure}

\begin{figure}[H]
\centering
\includegraphics[width=0.95\textwidth]{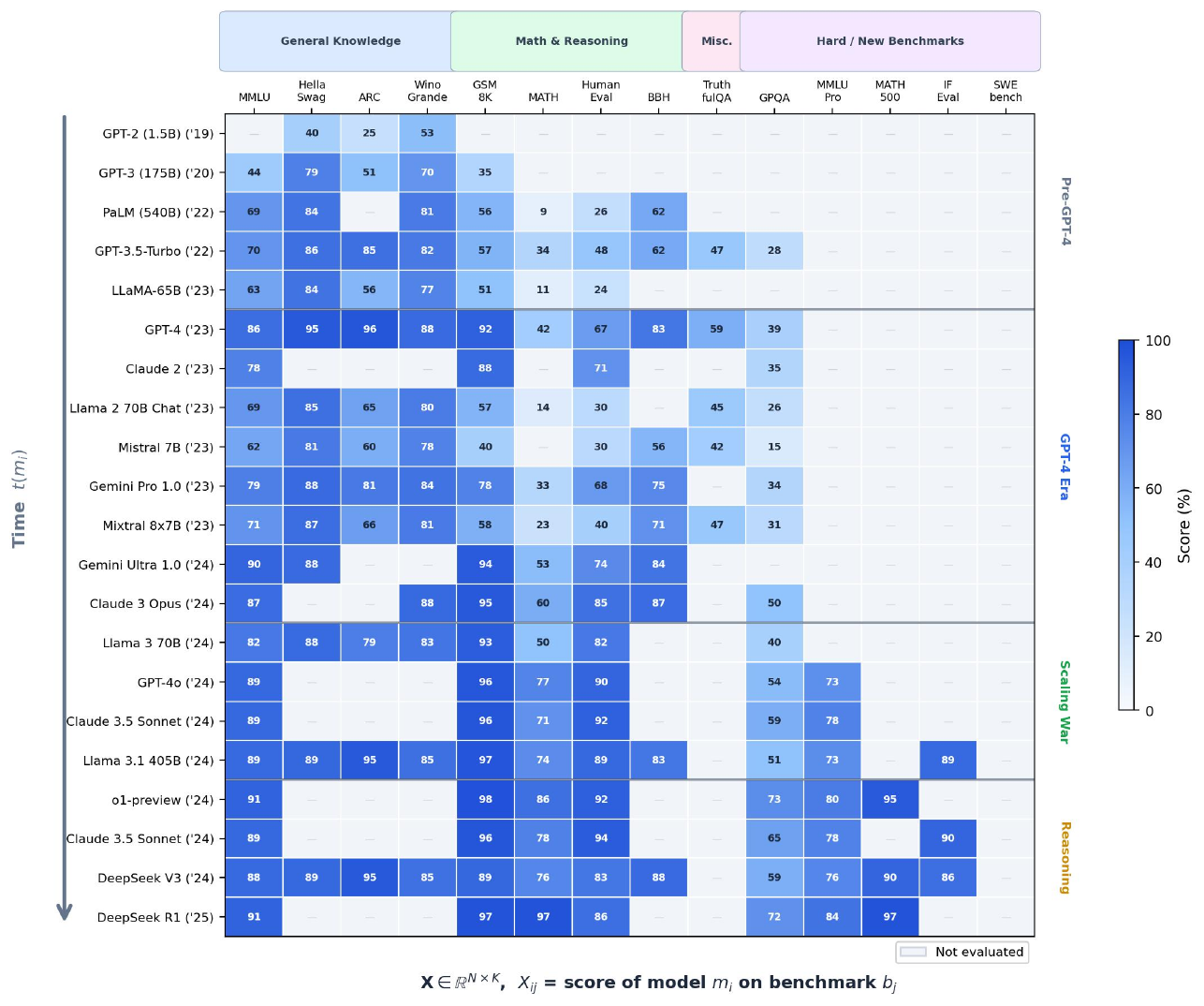}
\caption{\textbf{The score matrix $\mathbf{X}$.} A representative subset of 21 models (rows, ordered by release date) and 14 benchmarks (columns, grouped by cognitive domain). Cell color encodes score intensity (0--100\%); dashes mark missing evaluations. The matrix exhibits two key structural features: (i)~a gradient from low scores (upper-left) to high scores (lower-right), reflecting the correlated improvement that the PCA must dissect; and (ii)~a shifting pattern of sparsity, as early benchmarks (HellaSwag, ARC, WinoGrande) are not evaluated on late models, while late benchmarks (GPQA, MMLU-Pro, SWE-bench) do not exist for early models. Horizontal lines mark algorithmic-epoch boundaries; epoch labels are shown on the right.}
\label{fig:scorematrix}
\end{figure}

Let $\mathcal{M} = \{m_1, \ldots, m_N\}$ be a set of $N$ language models, each associated with a release date $t(m_i) \in \mathbb{R}$ and an organization $\text{org}(m_i)$. Let $\mathcal{B} = \{b_1, \ldots, b_K\}$ be a set of $K$ benchmarks. Define the \emph{score matrix}:
\begin{equation}
    \mathbf{X} \in \mathbb{R}^{N \times K}, \quad X_{ij} = \text{score of model } m_i \text{ on benchmark } b_j
\end{equation}
where scores are normalized to $[0,100]$ (percentage correct). Missing entries are denoted $X_{ij} = \texttt{NaN}$. The structure of the score matrix is shown schematically in Figure~\ref{fig:scorematrix}: models are ordered by release date (rows), benchmarks are grouped by cognitive domain (columns), and the pattern of missing data indicates sparsity in the upper-left (early models, early benchmarks) and lower-right (late models, new benchmarks).

\begin{definition}[Positive Manifold]
The score matrix $\mathbf{X}$ exhibits a \emph{positive manifold} if all pairwise Pearson correlations $r(b_j, b_k) > 0$ for $j \neq k$, computed on pairwise-complete observations with minimum sample size $n_{\min} \geq 5$.
\end{definition}

\begin{definition}[$G$-factor]
The $G$-factor is the first principal component of the standardized score matrix $\mathbf{Z} = \text{StandardScaler}(\mathbf{X})$. The \emph{$G$-loading} of benchmark $b_j$ is the $j$-th element of the first eigenvector $\mathbf{v}_1$ of the correlation matrix $\mathbf{R} = \mathbf{Z}^\top \mathbf{Z} / (N-1)$. The \emph{$G$-score} of model $m_i$ is its projection onto $\mathbf{v}_1$:
\begin{equation}
    G(m_i) = \mathbf{z}_i \cdot \mathbf{v}_1 = \sum_{j=1}^K z_{ij} v_{1j}
\end{equation}
\end{definition}

\subsection{Eigenvalue Diagnostics}

The eigenvalues $\lambda_1 \geq \lambda_2 \geq \cdots \geq \lambda_K$ of $\mathbf{R}$ encode the factor structure. Three diagnostic quantities characterize the factor structure:

\begin{enumerate}[label=(\roman*)]
    \item {Variance ratio:} $\rho_1 = \lambda_1 / \sum_k \lambda_k$. A high $\rho_1$ indicates a dominant general factor.
    \item {Dominance ratio:} $\delta = \lambda_1 / \lambda_2$. A high $\delta$ indicates that the first factor is clearly separated from the second.
    \item {Effective dimensionality:} $d_{\text{eff}} = \left(\sum_k \lambda_k\right)^2 / \sum_k \lambda_k^2$ (participation ratio). $d_{\text{eff}} \approx 1$ indicates a single dominant factor; $d_{\text{eff}} \approx K$ indicates uniform spread.
\end{enumerate}

For factor retention, the Kaiser criterion is applied ($\lambda_k > 1$) and Horn's parallel analysis \citep{horn1965}, which compares observed eigenvalues to those expected under random permutation of the data matrix.

\subsection{Connection to Psychometric $g$}

In psychometrics, the subjects are \emph{sampled from a natural population} (humans), and the positive manifold reflects shared cognitive architecture. In the LLM setting, the ``population'' of models is an engineered trajectory where each model is designed to improve upon its predecessors. This introduces some important structural differences. Models are not independent draws from a distribution but temporally ordered optimizations. This inflates between-era correlations (early models are bad at everything; late models are good at everything). All transformer-based LLMs are trained on overlapping internet-scale corpora, potentially inducing positive correlations through shared data rather than shared mechanism. Benchmarks are designed with knowledge of model capabilities, and labs optimize against known benchmarks, creating a feedback loop absent from psychometric testing. 

\section{ Results}

\subsection{Data}

The score matrix comprises $N=39$ models and $K=14$ benchmarks spanning February 2019 to December 2025. Models represent major releases from OpenAI, Anthropic, Google, Meta, DeepSeek, and Mistral. Benchmark scores are drawn from published technical reports, model cards, and third-party evaluations \citep{epochai2024}. All scores are converted to a 0--100 percentage scale. The matrix has 42\% overall coverage, with higher coverage for earlier benchmarks (MMLU: 77\%) and lower for newer ones (SWE-bench Verified: 21\%). The problem with the current state of data is its small sample size. This makes rigorous statistical conclusions provisional. When statistical confidence is significant this will be stated clearly in the manuscript. Otherwise the results should be seen as descriptive and awaiting further data. See Appendix \ref{app:parallel} for permutation analysis. 

\subsection{Benchmark Saturation}

\begin{figure}[H]
\centering
\includegraphics[width=0.92\textwidth]{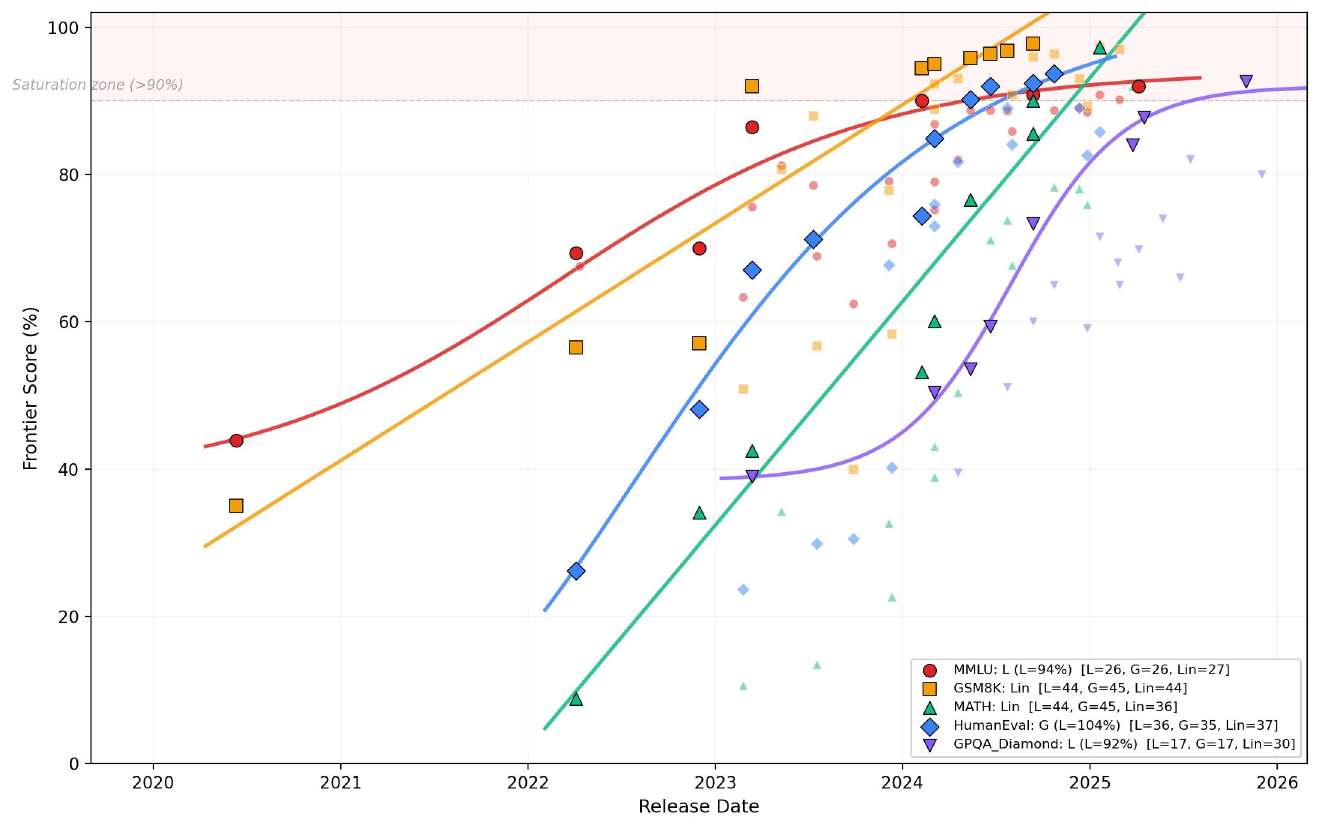}
\caption{\textbf{The phenomenon to be dissected: benchmark performance rising across all tasks simultaneously.} Bold markers show the running-maximum score (frontier envelope) for each benchmark; faded markers show all individual model scores. Curves are the best-fit growth model selected by AIC (Table~\ref{tab:saturation}). MMLU is best fit by a logistic with asymptote $L = 94\%$; GPQA Diamond by a logistic with $L = 92\%$; HumanEval by a Gompertz with $L = 104\%$ (no saturation yet). The shaded region marks $>90\%$ scores where discriminability is lost. The correlated rise across all five benchmarks, despite their testing different cognitive demands, is the positive manifold in its raw form.}
\label{fig:saturation}
\end{figure}

The raw data exhibit a clear pattern: frontier model performance on every major benchmark has increased monotonically over the 2020--2025 period, with different benchmarks reaching saturation ($>90\%$) at different times (Figure~\ref{fig:saturation}). This represents correlated growth across diverse tasks including general knowledge (MMLU), mathematical reasoning (MATH, GSM8K), code generation (HumanEval), and scientific knowledge (GPQA Diamond). 

To characterize the growth dynamics, three functional forms are fit to the \emph{frontier envelope}---the running maximum score across all models at each time point---for each benchmark: 4-parameter logistic, 4-parameter Gompertz, and linear, selecting by AIC (Table~\ref{tab:saturation}).

\begin{table}[H]
\centering
\caption{\textbf{Growth model fits to the frontier envelope.} AIC values for logistic (L), Gompertz (G), and linear (Lin) models fitted to the running-maximum score trajectory. Bold indicates the selected model. $L$ denotes the estimated asymptote for sigmoidal fits; $n$ is the number of frontier-envelope points.}
\label{tab:saturation}
\begin{tabular}{lcccccl}
\toprule
\textbf{Benchmark} & $n$ & \textbf{AIC$_\text{L}$} & \textbf{AIC$_\text{G}$} & \textbf{AIC$_\text{Lin}$} & \textbf{Best} & $L$ \\
\midrule
MMLU & 7 & \textbf{25.9} & 26.1 & 27.0 & Logistic & 94\% \\
GSM8K & 10 & 44.3 & 45.0 & \textbf{43.5} & Linear & --- \\
MATH & 9 & 43.5 & 44.8 & \textbf{36.0} & Linear & --- \\
HumanEval & 10 & 35.8 & \textbf{34.7} & 37.0 & Gompertz & 104\% \\
GPQA Diamond & 8 & \textbf{16.7} & 16.8 & 29.6 & Logistic & 92\% \\
\bottomrule
\end{tabular}
\end{table}

A benchmark saturates when frontier models approach its ceiling, eliminating variance and rendering the benchmark useless for factor analysis. Define the \emph{discriminability} of benchmark $b_j$ at time $t$ as the standard deviation of scores among contemporaneous models: $D_j(t) = \text{SD}(\{X_{ij} : |t(m_i) - t| < \Delta\})$. When $D_j(t) \to 0$, benchmark $b_j$ is saturated. This creates a \emph{moving battery}: the set of informative benchmarks changes over time, complicating longitudinal comparison.

\subsection{The Positive Manifold}

\begin{figure}[H]
\centering
\includegraphics[width=0.72\textwidth]{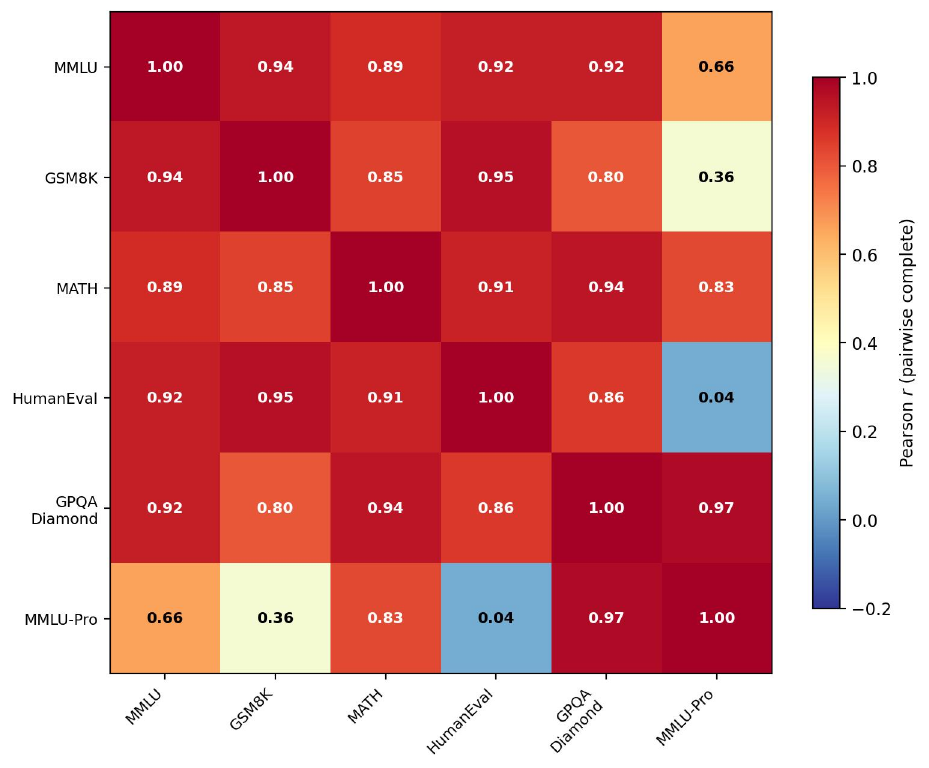}
\caption{\textbf{The positive manifold in LLM benchmarks.} Pairwise Pearson correlations across six major benchmarks, computed using pairwise-complete observations (minimum $n=5$ pairs per cell). All 15 off-diagonal entries are positive, satisfying the Spearman criterion for a general factor. Note the near-zero correlation between HumanEval and MMLU-Pro ($r=0.04$), suggesting that code generation and hard general knowledge are nearly independent.}
\label{fig:manifold}
\end{figure}

Across the six benchmarks with sufficient pairwise coverage (MMLU, GSM8K, MATH, HumanEval, GPQA Diamond, MMLU-Pro), all 28 pairwise correlations are positive, confirming the positive manifold (Figure~\ref{fig:manifold}). With the updated data matrix, robust pairwise correlations ($n \geq 5$) can be computed across 8 benchmarks (MMLU, HellaSwag, ARC, WinoGrande, GSM8K, MATH, HumanEval, BBH). Correlations range from $r = 0.42$ (WinoGrande $\times$ MATH) to $r = 0.96$ (BBH $\times$ MMLU; BBH $\times$ GSM8K), with a mean of $\bar{r} = 0.82$.

\subsection{Factor Structure}

PCA on the five-benchmark core battery (MMLU, GSM8K, MATH, HumanEval, GPQA Diamond) for the 19 models with complete data yields the eigenvalue structure shown in Table~\ref{tab:eigenvalues}. Only PC1 exceeds the Kaiser criterion ($\lambda > 1$), supporting a single-factor solution. PC1 explains 90.0\% of total variance---substantially stronger than the 40--60\% typically attributed to $g$ in human psychometrics, and consistent with the 66\% reported by Ili\'c and Gignac \citep{ilic2024evidence} on a much larger sample of 591 models.
\begin{table}[H]
\centering
\caption{\textbf{Eigenvalue decomposition of the 5-benchmark core battery.} PCA on standardized scores for 19 models with complete data across MMLU, GSM8K, MATH, HumanEval, and GPQA Diamond. Only PC1 exceeds the Kaiser criterion ($\lambda > 1$).}
\label{tab:eigenvalues}
\begin{tabular}{cccc}
\toprule
\textbf{Component} & $\lambda$ & \textbf{Variance Explained} & \textbf{Cumulative} \\
\midrule
PC1 ($G$) & 4.50 & 90.0\% & 90.0\% \\
PC2 & 0.35 & 6.6\% & 96.6\% \\
PC3 & 0.10 & 1.9\% & 98.5\% \\
PC4 & 0.05 & 0.9\% & 99.4\% \\
PC5 & 0.03 & 0.6\% & 100.0\% \\
\bottomrule
\end{tabular}
\end{table}

\begin{figure}[H]
\centering
\includegraphics[width=0.55\textwidth]{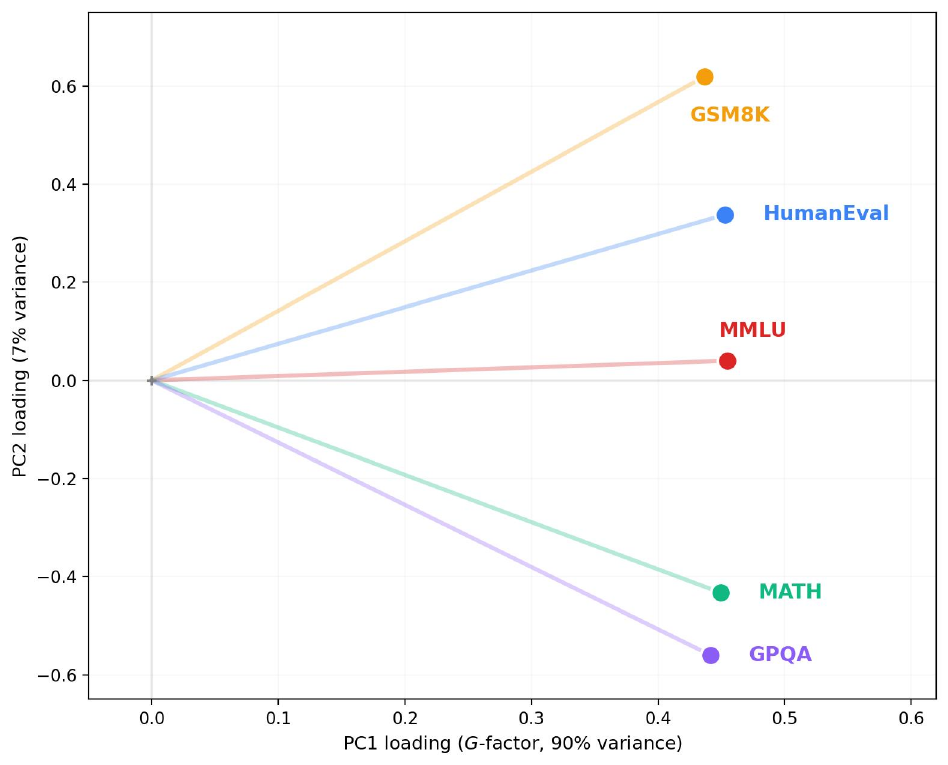}
\caption{\textbf{Factor loading plot for the 5-benchmark core battery.} Arrows show each benchmark's loading on PC1 ($G$-factor, 90\% variance) and PC2 (7\% variance). All benchmarks load positively on PC1, confirming a general factor. PC2 separates an execution/fluency pole (GSM8K, HumanEval---positive PC2) from a reasoning pole (MATH, GPQA---negative PC2). MMLU is near the origin on PC2, contributing primarily to $G$ rather than to the residual structure.}
\label{fig:loadings}
\end{figure}

All benchmarks load positively and near-uniformly on PC1 (range: $+0.44$ to $+0.46$), with the highest loadings on MMLU ($+0.46$) and HumanEval ($+0.45$) (Figure~\ref{fig:loadings}). PC2 (7\% variance) separates an execution cluster (GSM8K at $+0.62$, HumanEval at $+0.34$: positive PC2 loadings) from a reasoning cluster (GPQA at $-0.56$, MATH at $-0.43$: negative PC2 loadings), with MMLU near zero ($+0.04$). 

\subsection{$G$-Scores Across Models and Time}

\begin{figure}[H]
\centering
\includegraphics[width=0.85\textwidth]{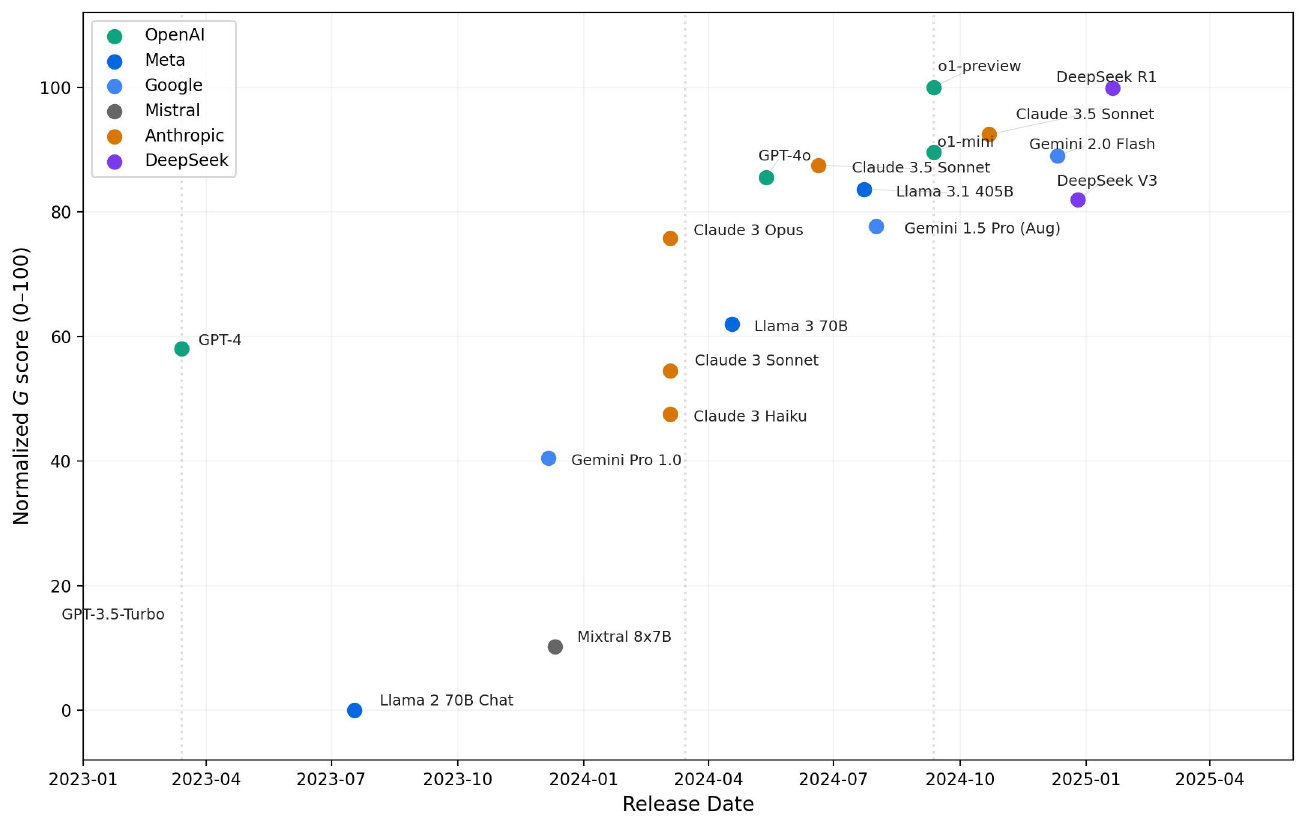}
\caption{\textbf{Normalized $G$: LLM general intelligence factor over time.} Each point represents a model's projection onto PC1 of the 5-benchmark battery (MMLU, GSM8K, MATH, HumanEval, GPQA Diamond), rescaled to a 0--100 range where 0 is the lowest-scoring model (Llama~2~70B~Chat) and 100 is the highest (o1-preview). PC1 captures 90\% of total variance across the battery. The 19 models with complete data on all five benchmarks are shown, colored by organization. The normalization makes $G$ directly interpretable as a percentile-like score: a model at $G = 50$ is halfway between the least and most capable systems in the sample.}
\label{fig:timeline}
\end{figure}

Projecting each model onto PC1 and normalizing to a 0--100 scale yields a $G$-score (Figure~\ref{fig:timeline}). $G$ increases monotonically with release date, rising from Llama~2~70B~Chat ($G = 0$, July 2023, the lowest-scoring model in the complete-data set) through GPT-4 ($G = 58$) and GPT-4o ($G = 86$) to the post-September 2024 models (o1-preview, $G = 100$; DeepSeek~R1, $G = 100$). The expanded sample now includes models from GPT-3.5-Turbo ($G = 14$) through the Gemini and Claude families, providing a denser trajectory. The rate of $G$-growth accelerates around mid-2024, coinciding with inference-time reasoning.

\section{Temporal Decomposition}

\subsection{Epoch Structure}

The model timeline divides into \emph{algorithmic epochs}---periods defined by the dominant paradigm of model development:

\begin{center}
\begin{tabular}{llp{6cm}}
\toprule
\textbf{Epoch} & \textbf{Period} & \textbf{Defining innovation} \\
\midrule
Epoch I & 2019.02 -- 2023.02 & Scaling laws, basic RLHF \\
Epoch II & 2023.03 -- 2024.03 & Frontier-scale dense models \\
Epoch III & 2024.04 -- 2024.09 & Multi-lab competition, MoE \\
Epoch IV & 2024.09 -- 2025+ & Chain-of-thought, inference-time compute \\
\bottomrule
\end{tabular}
\end{center}

\subsection{Epoch-Resolved Factor Structure}

\begin{figure}[H]
\centering
\includegraphics[width=0.85\textwidth]{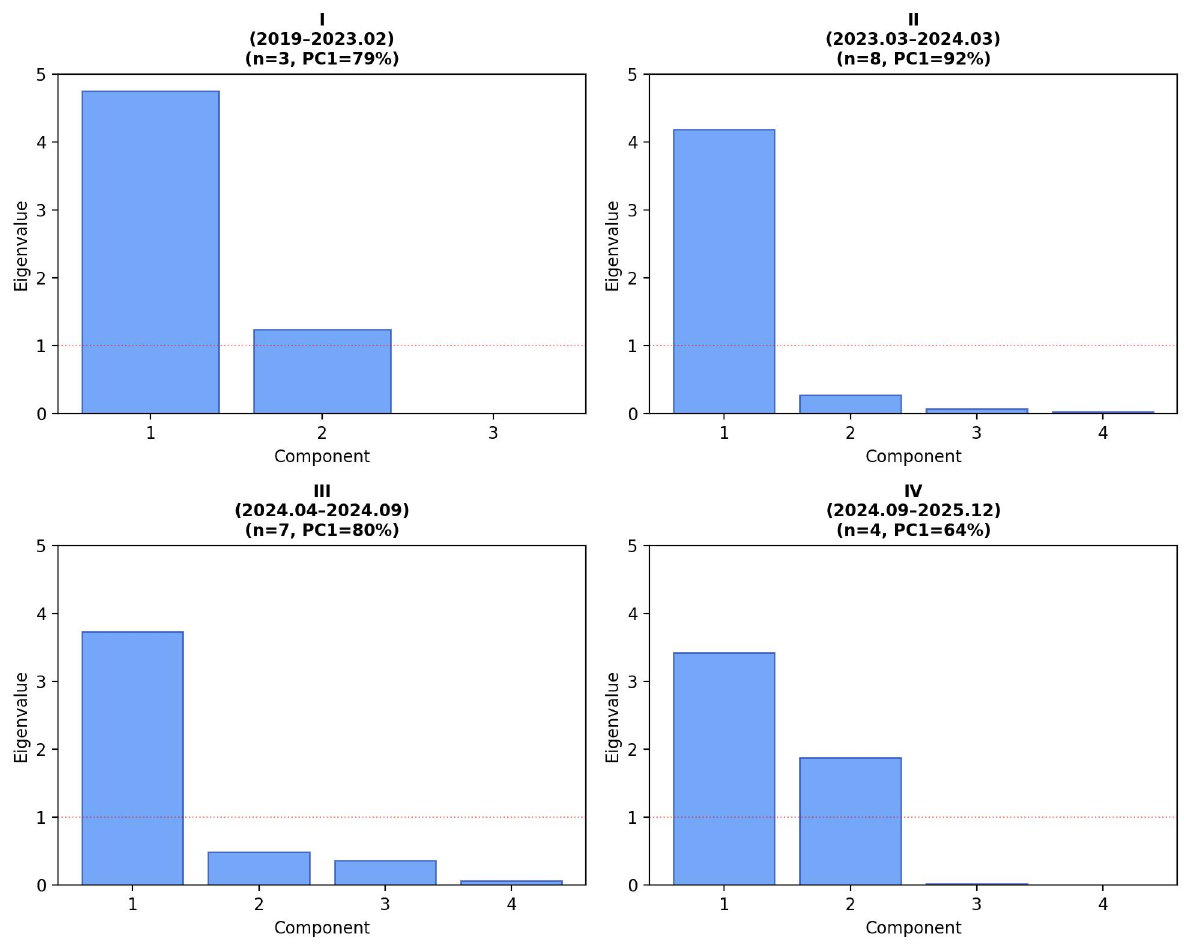}
\caption{\textbf{Scree plots across algorithmic epochs.} Eigenvalue decomposition of the 4-benchmark battery (MMLU, GSM8K, MATH, HumanEval) computed within each epoch. The red dashed line marks the Kaiser criterion ($\lambda = 1$). During Epoch~II (2023--2024.03), a single dominant factor captures 92\% of variance and no second eigenvalue approaches 1.0. In Epoch~IV (2024.09+), a second eigenvalue (1.88) increases alongside the creation of inference-time reasoning models (o1, DeepSeek~R1).}
\label{fig:scree}
\end{figure}

The scree plots across epochs (Figure~\ref{fig:scree}) reveal the predicted structure. The key quantities are summarized in Table~\ref{tab:epochs}.

\begin{table}[H]
\centering
\caption{\textbf{Factor structure across algorithmic epochs.} PCA on a 4-benchmark battery (MMLU, GSM8K, MATH, HumanEval) within each epoch. $\rho_1$ is the variance explained by PC1; $\lambda_2$ is the second eigenvalue. The Kaiser criterion ($\lambda_2 > 1$) is exceeded only in Epochs~I and IV, suggestive of a two-factor structure but with insufficient data to statistically validate.}
\label{tab:epochs}
\begin{tabular}{lcccc}
\toprule
\textbf{Epoch} & $n$ & \textbf{Mean} $r$ & $\rho_1$ & $\lambda_2$ \\
\midrule
I (2019--2023) & 3 & 0.72 & 79\% & 1.24 \\
II (2023--2024.03) & 8 & 0.90 & 92\% & 0.28 \\
III (2024.04--2024.09) & 7 & 0.73 & 80\% & 0.49 \\
IV (2024.09--2025) & 4 & 0.43 & 64\% & 1.88 \\
\bottomrule
\end{tabular}
\end{table}

The $G$-factor peaks during Epoch~II (2023--2024.03), when pure scaling dominates and all labs improve uniformly across tasks. It then splinters in Epoch~IV (2024.09+), where $\lambda_2$ rises above 1.0---suggestive of a two-factor structure, though with $n = 4$ this cannot be confirmed statistically (see Appendix~\ref{app:parallel}). This second factor appears to distinguish ``depth of search'' (MATH, GPQA: reasoning-chain models excel) from ``breadth of recall'' (MMLU, GSM8K, HumanEval: scaling models excel). The dominance ratio $\delta = \lambda_1 / \lambda_2$ drops from 15:1 (Epoch~II) to 1.8:1 (Epoch~IV), a pronounced structural break.

\subsection{Expanding-Window Dynamics}

\begin{figure}[H]
\centering
\includegraphics[width=0.85\textwidth]{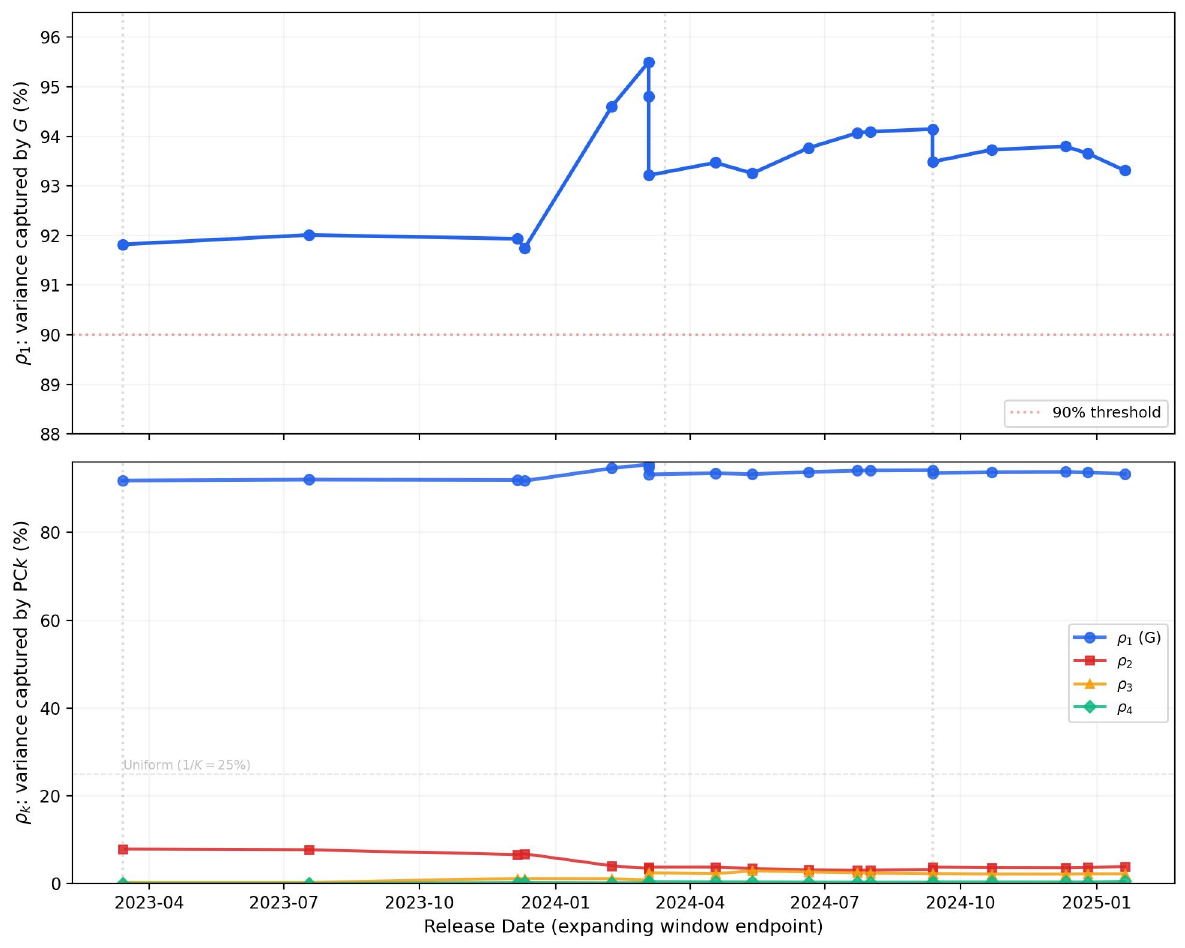}
\caption{\textbf{Expanding-window tracking changing $G$}. Models are added in chronological order of release date; at each step, PCA is computed on the 4-benchmark battery. \textit{Top:}~$\rho_1$, the fraction of total variance captured by PC1 ($G$). The 90\% threshold (red dotted line) is exceeded throughout, with a peak at 95.5\% around the Claude~3~Opus release (early 2024). \textit{Bottom:}~All four normalized variance fractions $\rho_1, \rho_2, \rho_3, \rho_4$ plotted on the same scale. $G$ (blue) dominates throughout; $\rho_2$ (red) is the only component that shows any temporal dynamics, growing modestly as post-2024.09 models enter the window. The dashed gray line at 25\% marks the uniform distribution ($\rho_k = 1/K$, no factor structure). }
\label{fig:expanding}
\end{figure}

An expanding-window analysis, produced by adding models chronologically and recomputing the eigenvalue structure at each step, reveals elements of the dynamics of $G$ (Figure~\ref{fig:expanding}). The top panel tracks $\rho_1$, the normalized fraction of total variance captured by PC1, which remains consistently above 90\% from the earliest window through the entire timeline, peaking at 95.5\% around the Claude~3~Opus release (early 2024) and settling to 93.3\% by the final window. The bottom panel shows all four normalized variance fractions $\rho_k$ on a common scale, $\rho_1$ occupies the upper reaches of the plot while $\rho_2$, $\rho_3$, and $\rho_4$ are compressed near the floor. The ratio $\rho_1/\rho_2$ peaks at 31:1 around the Llama~3.1 release and then declines to 24:1 as post-2024.09 models enter the window. This decline in the ratio occurs not because $\rho_1$ collapses but because $\rho_2$ grows from 3.1\% to 3.9\%---a subtle redistribution of variance from the general factor to the residual structure. The reference line at 25\% (uniform distribution across $K=4$ components) makes clear how far the empirical spectrum is from ``no structure'': even at its weakest, $G$ captures nearly four times the variance of the next component.

\subsection{Does the Positive Manifold Reduce in Dimension?}

\begin{figure}[H]
\centering
\includegraphics[width=0.92\textwidth]{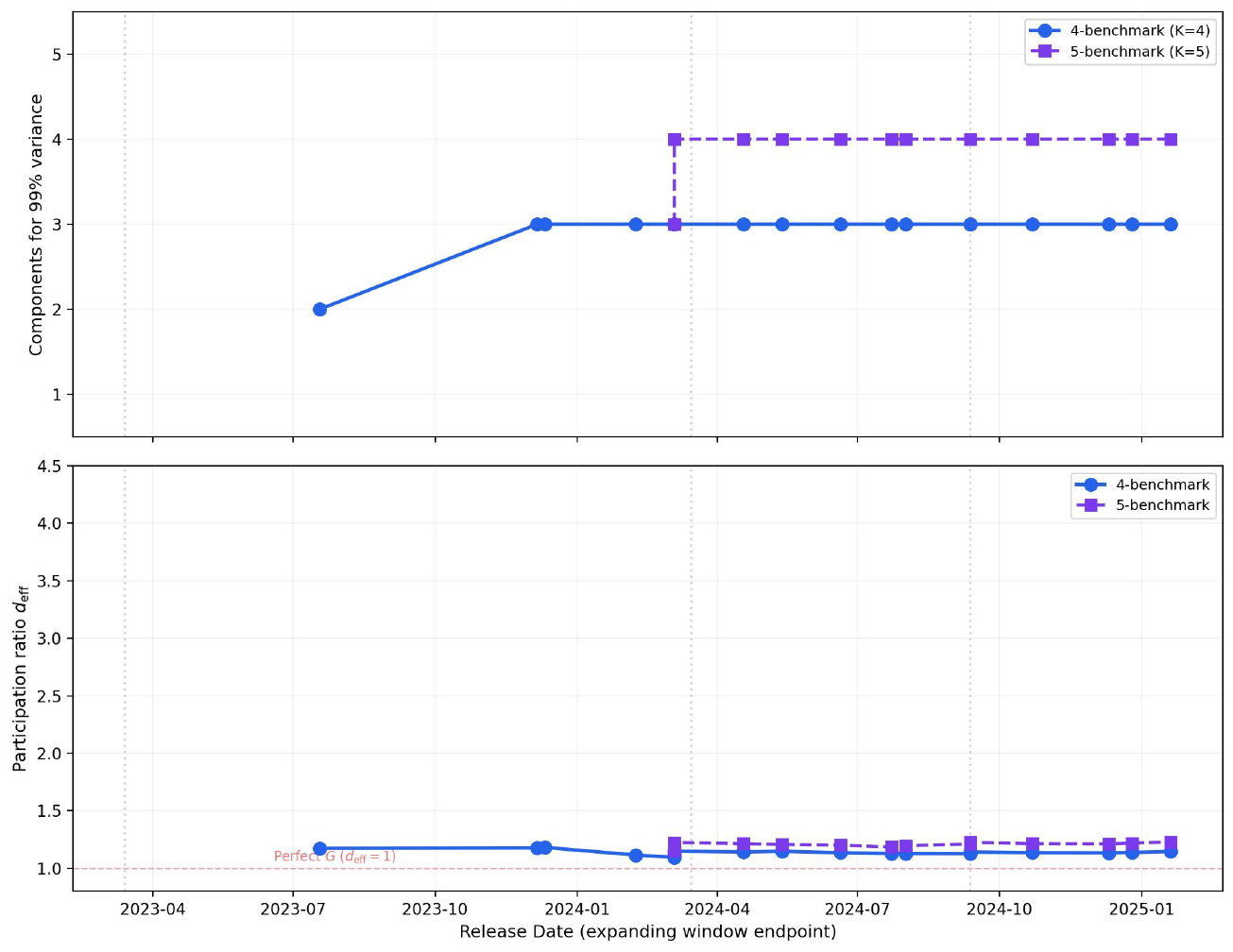}
\caption{\textbf{Effective dimensionality of the LLM benchmark space through time.} Expanding-window analysis on two benchmark batteries. \textit{Top:}~Number of principal components required for 99\% cumulative variance. On the 4-benchmark battery (blue circles), dimensionality stabilizes at 3 of 4---near-maximal compression. On the 5-benchmark battery (purple squares, adding GPQA Diamond), dimensionality rises from 3 to 5 as post-2024.09 models enter the window. \textit{Bottom:}~Participation ratio $d_{\text{eff}}$, a continuous measure ranging from 1 (perfect single-factor structure) to $K$ (no structure). The 4-benchmark battery remains near $d_{\text{eff}} \approx 1.1$ throughout---a near-perfect $G$. The 5-benchmark battery rises from 1.3 to 1.9, suggestive of a second factor associated with inference-time reasoning. Vertical dotted lines mark algorithmic-epoch boundaries.}
\label{fig:effdim}
\end{figure}

A strict positive manifold implies that the effective dimensionality of the benchmark space should decrease as $G$ increases. Two measures track this through the expanding window: (i)~the number of principal components required to capture 99\% of total variance, and (ii)~the participation ratio $d_{\text{eff}} = (\sum_k \lambda_k)^2 / \sum_k \lambda_k^2$, which equals 1 when a single factor dominates and $K$ when variance is uniformly spread (Figure~\ref{fig:effdim}).

On the \textbf{4-benchmark battery} (MMLU, GSM8K, MATH, HumanEval), the participation ratio remains close to its minimum at $d_{\text{eff}} \approx 1.1$--$1.2$ throughout the entire timeline. $G$ has already absorbed essentially all the variance; the manifold is as compressed as it can be. Three of four components are needed for 99\% variance, but this residual dimensionality reflects only minor task-specific variance.

On the \textbf{5-benchmark battery} (adding GPQA Diamond), a different picture emerges. The participation ratio begins at $d_{\text{eff}} = 1.3$ when GPQA first enters the window (mid-2024) and rises to $d_{\text{eff}} = 1.9$ as reasoning models arrive---a 40\% increase in effective dimensionality. The number of components for 99\% variance rises from 3 to 5 (i.e., all components).

Within a fixed set of models the manifold remains maximally compressed: $G$ captures nearly everything, and adding models does not alter the dimensionality. But when a new capability axis appears that the existing battery was not designed to measure the effective dimensionality increases. The manifold shows evidence of growing along a new dimension that harder benchmarks reveal.  Each algorithmic epoch compresses the space within its own test battery and epoch transitions open new dimensions that require new benchmarks to detect. The positive manifold holds within each epoch, but the dimensionality of the full manifold grows as the set of distinguishable capabilities expands.

\begin{figure}[H]
\centering
\includegraphics[width=0.95\textwidth]{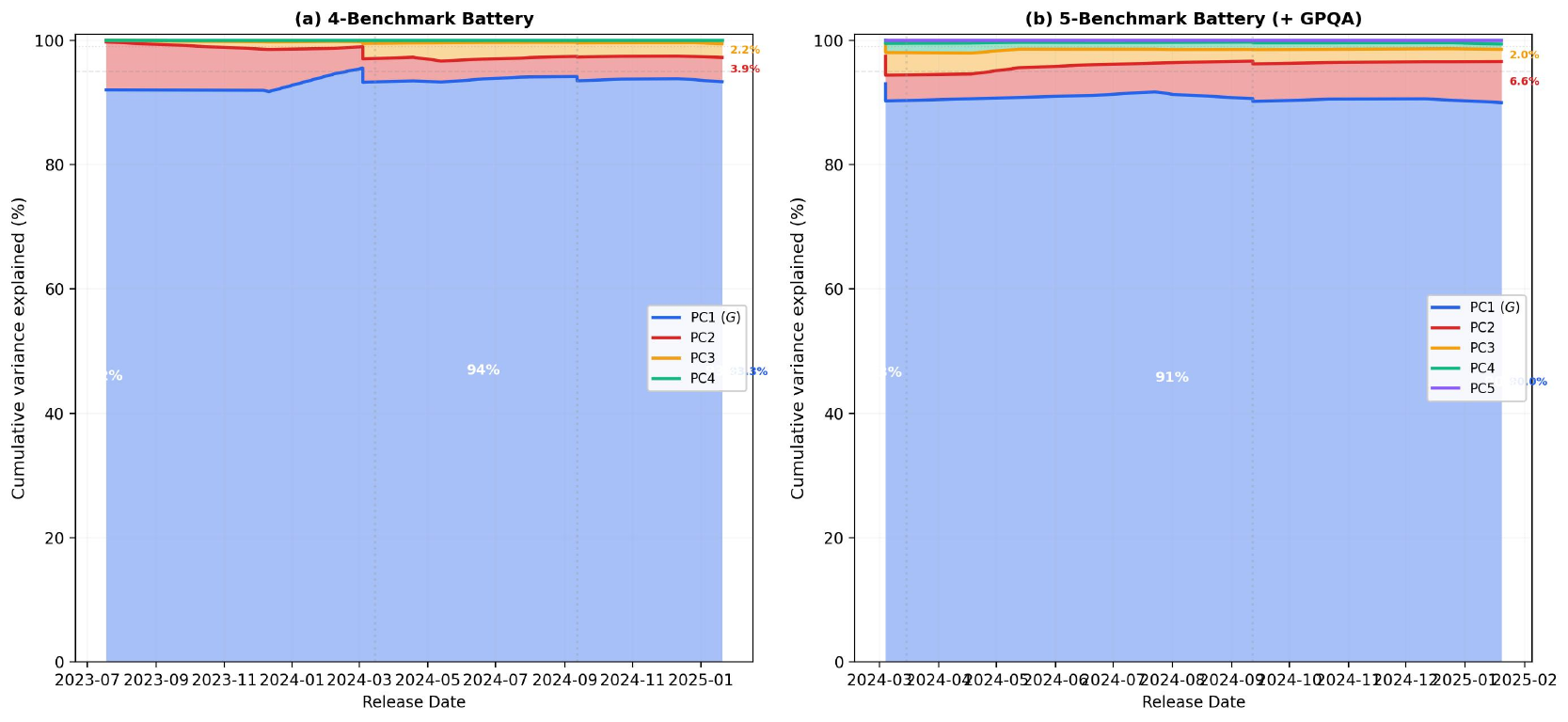}
\caption{\textbf{Eigenvalue spectrum through time: cumulative variance partitioned by principal component.} Each colored band shows one component's marginal contribution to cumulative variance in the expanding-window PCA; the top of each band is the cumulative variance through that component. (a)~4-benchmark battery: PC1 ($G$, blue) accounts for 92--95\% throughout; PC1$+$PC2 exceed 97\% at every time point. The spectrum is effectively one-dimensional---a near-perfect general factor. (b)~5-benchmark battery (adding GPQA Diamond): the spectrum visibly unfolds as post-2024.09 models enter the window. With the expanded sample ($n = 19$), for the 5-benchmark spectrum PC1 remains near 90\% throughout. }
\label{fig:spectrum}
\end{figure}

The full eigenvalue spectrum through time makes this conjecture more visible (Figure~\ref{fig:spectrum}). Each component's marginal contribution to cumulative variance is shown as a stacked band, so the vertical extent of the blue band ($G$) relative to the total directly encodes the strength of the general factor.

On the 4-benchmark battery (panel~a), PC1 accounts for 92--95\% of variance throughout, and PC1$+$PC2 together exceed 97\% at every time point. The higher components (PC3, PC4) contribute only thin slivers and the eigenvalue spectrum is essentially one-dimensional from the very first window. This is a near-maximal $G$ in which one dimension suffices to reconstruct the entire benchmark space to within a few percent.

On the 5-benchmark battery (panel~b), the spectrum undergoes a degree of structural transition. Initially, PC1 explains 85\% and PC1$+$PC2 reach 97\%---comparable to the 4-benchmark picture. With the expanded GPQA coverage, the 5-benchmark expanding window now begins earlier (from Llama~2~70B~Chat onward) and includes 19 models. PC1 remains above 90\% throughout most of the trajectory, settling at 90\% by the final window. PC2 accounts for only 7\% and PC3 for 2\%. 

\subsection{Eigenvalue Change-Point Analysis}

The CUSUM (cumulative sum) statistic, applied to the sequences $\{\rho_1(t)\}$ and $\{\delta(t)\}$ generated by the expanding-window PCA on the 4-benchmark battery. Under the null hypothesis of a stationary factor structure, the CUSUM statistic fluctuates randomly around zero; a systematic excursion indicates a change in the mean level of the diagnostic. Significance follows from a permutation test (10,000 random reorderings of the sequence).

\begin{figure}[H]
\centering
\includegraphics[width=0.92\textwidth]{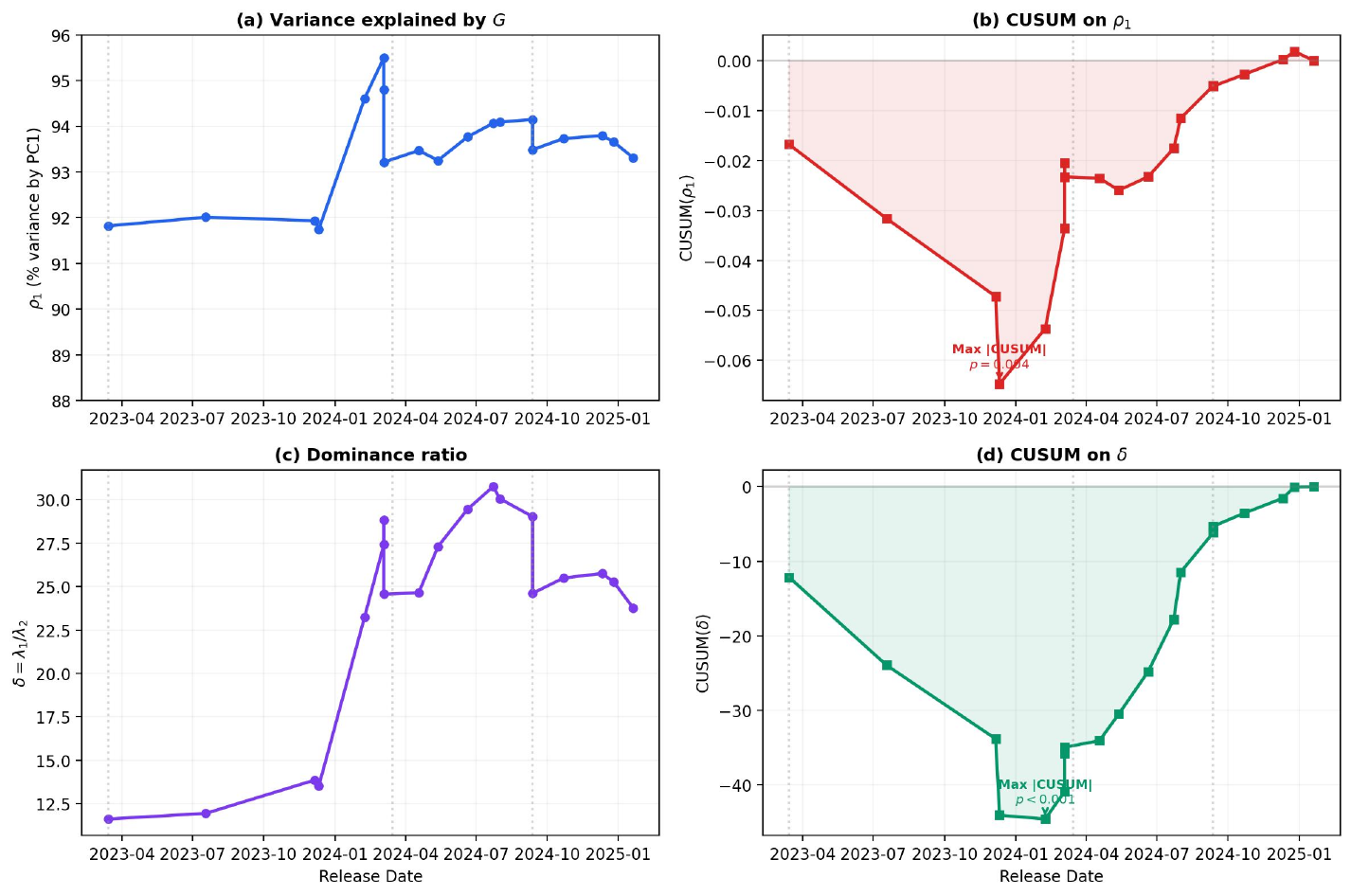}
\caption{\textbf{Test~(i): CUSUM change-point analysis on eigenvalue diagnostics.} (a)~Variance explained by PC1 ($\rho_1$) in the expanding-window 4-benchmark PCA, showing a peak around the Claude~3~Opus/Gemini~Ultra releases (early 2024) followed by decline. (b)~CUSUM statistic on $\rho_1$, with maximum absolute deviation marked; $p = 0.004$ by permutation test, confirming a significant structural break. (c)~Dominance ratio $\delta = \lambda_1/\lambda_2$, showing a sharp jump from $\sim$12 to $\sim$28 when the GPT-4-era models enter, followed by gradual decline. (d)~CUSUM on $\delta$; $p < 0.001$, the strongest signal. Vertical dotted lines mark epoch boundaries. Both CUSUM statistics identify the early-2024 transition as a significant change point in the factor structure.}
\label{fig:cusum}
\end{figure}

Both diagnostics yield significant change points (Figure~\ref{fig:cusum}). The CUSUM on $\rho_1$ reaches its maximum deviation in early 2024 ($p = 0.004$, permutation test), coinciding precisely with the transition from Epoch~II to Epoch~III (early 2024). The CUSUM on $\delta = \lambda_1/\lambda_2$ is even more decisive ($p < 0.001$), with the maximum deviation occurring at the same boundary. A caveat is in order: the expanding-window eigenvalue sequence is not a standard time series, since successive values share $n - 1$ of $n$ data points, and the permutation test (which shuffles model order) conflates temporal trend with structural change.  The pattern suggests that $G$ peaks in early 2024 and then declines, not because $G$ disappears, but because the second factor increases.

\subsection{Eigenvector Rotation}

\begin{figure}[H]
\centering
\includegraphics[width=0.92\textwidth]{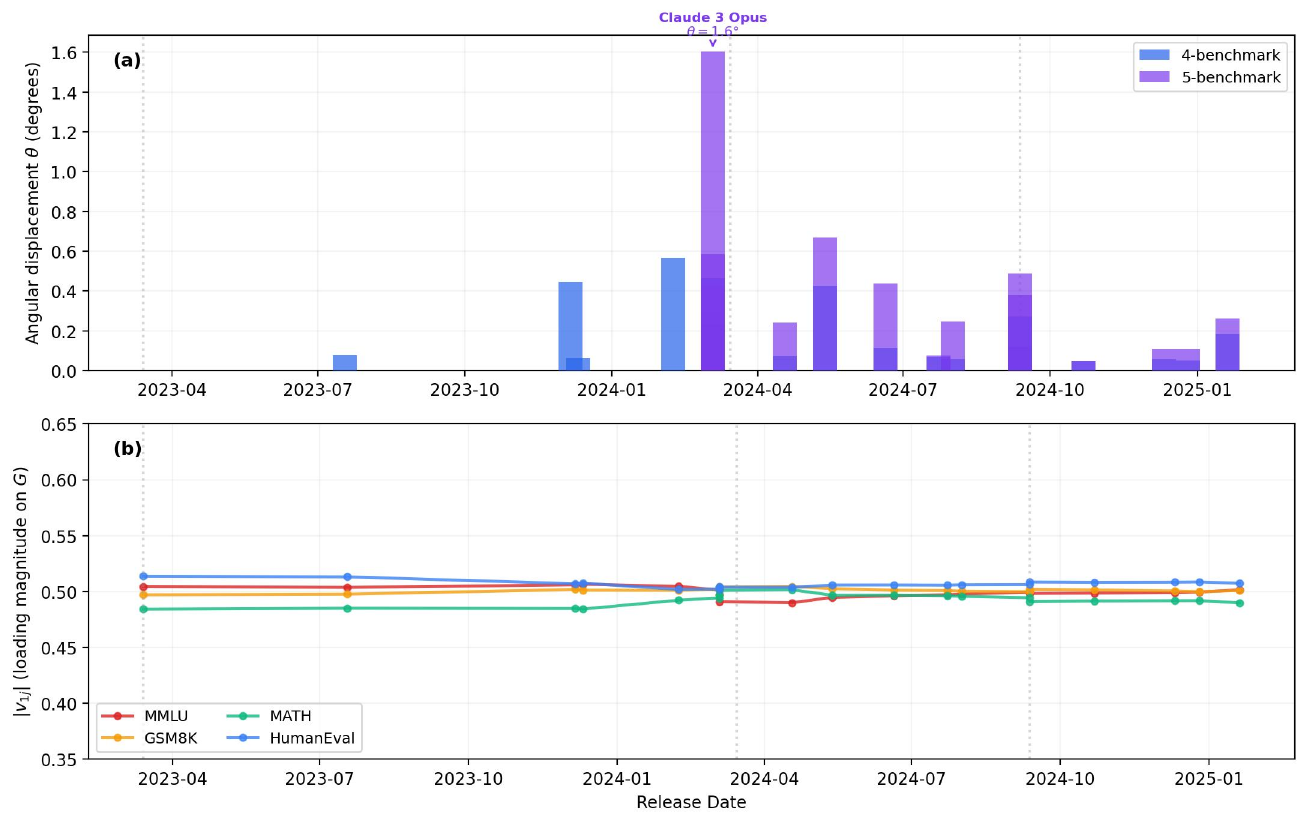}
\caption{\textbf{Test~(ii): Eigenvector alignment.} (a)~Angular displacement $\theta$ between consecutive first eigenvectors as each model enters the expanding window. Blue bars: 4-benchmark battery (max $\theta = 0.57°$, near-perfect stability). Purple bars: 5-benchmark battery (max $\theta = 6.4°$ at DeepSeek~V3 entry). The 5-benchmark battery reveals rotations invisible to the 4-benchmark analysis because GPQA Diamond introduces a dimension along which post-2024.09 models diverge from pre-2024.09 models. (b)~Individual benchmark loadings $|v_{1j}|$ on $G$ over time (4-benchmark battery), confirming structural invariance: all four loadings remain within a 0.04-unit band across the entire timeline.}
\label{fig:alignment}
\end{figure}

For consecutive expanding windows $W_t$ and $W_{t+1}$, the cosine similarity is computed $\cos\theta$ between first eigenvectors and report the angular displacement $\theta$ in degrees (Figure~\ref{fig:alignment}).

On the 4-benchmark battery the maximum angular displacement across all 18 steps is $0.57°$. The calculation of $G$, or which benchmarks it weights, does not change as models are added. This near-perfect alignment indicates that, within the 4-benchmark space, the general factor is structurally invariant across the entire 2023--2025 timeline.

On the 5-benchmark battery, a different pattern emerges. Most steps show small rotations ($1°$--$3°$), but the entry of DeepSeek~V3 produces a $6.4°$ rotation---an order of magnitude larger than anything observed in the 4-benchmark analysis. This is a rotation of the $G$-factor following the addition of a model with a distinctive reasoning/knowledge profile thereby changing the loadings across benchmarks. The lower panel of Figure~\ref{fig:alignment} confirms that individual benchmark loadings on $G$ are essentially invariant in the 4-benchmark battery (all between $|v_{1j}| = 0.48$ and $0.52$ throughout), consistent with a fixed general factor.

\paragraph{Leave-one-out Validation} To calibrate the magnitude of the $6.4°$ rotation, a leave-one-out (LOO) analysis drops each model in turn from the complete-case set, recomputes PCA on the remaining $n-1$ models, and measures the angular displacement of PC1 from the full-sample eigenvector. This establishes a baseline distribution of single-model influence on the direction of $G$.

On the 4-benchmark battery ($n = 22$), the LOO perturbations are negligible: mean $\theta_{\text{LOO}} = 0.08°$, max $= 0.19°$ (dropping PaLM~540B), 95th percentile $= 0.18°$. No single model perturbs the eigenvector by more than $0.2°$. The general factor in this battery is structurally invariant to any individual model.

On the 5-benchmark battery ($n = 19$), the picture is different and more revealing. The LOO displacements are substantially larger: mean $\theta_{\text{LOO}} = 1.9°$, with a range from $0.28°$ (Claude~3~Opus) to $5.58°$ (DeepSeek~V3). Three models produce rotations exceeding $3°$: DeepSeek~V3 ($5.58°$), GPT-4 ($3.35°$), and DeepSeek~R1 ($3.16°$). The $6.4°$ rotation observed when DeepSeek~V3 \emph{enters} the expanding window exceeds the LOO maximum of $5.58°$ from \emph{dropping} it, but only by a factor of 1.15. DeepSeek~V3 is the single most influential model in the 5-benchmark battery; adding or removing it produces comparable perturbations.

\subsection{Test (iii): Partial Correlation Structure After Removing $G$}

\begin{figure}[H]
\centering
\includegraphics[width=0.95\textwidth]{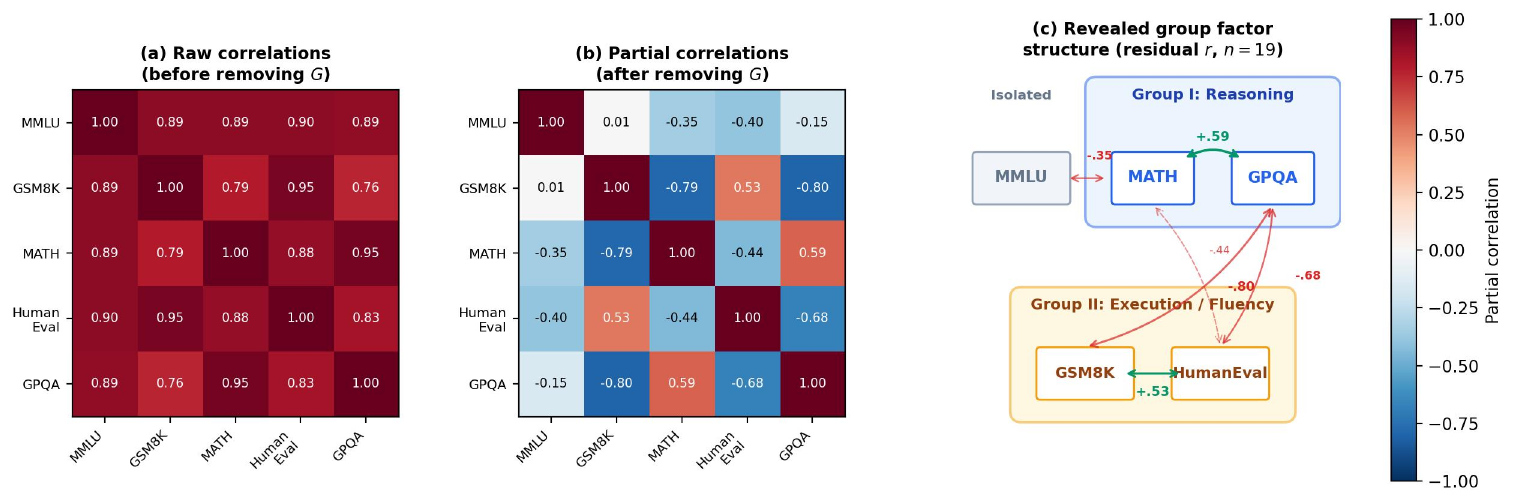}
\caption{\textbf{Partial correlation structure after removing $G$.} (a)~Raw correlation matrix (5-benchmark battery, $n = 19$ models with complete data): all 10 off-diagonal correlations positive. (b)~Partial correlation matrix after projecting out PC1: 7 of 10 correlations are now negative, with mean $r_{\text{resid}} = -0.24$. The positive manifold is entirely attributable to $G$. (c)~Revealed group factor structure derived from the residual correlations. Group~I (Reasoning: MATH, GPQA) and Group~II (Execution/Fluency: GSM8K, HumanEval) show positive within-group and negative between-group residual correlations; MMLU is isolated, with near-zero or negative residual correlations with all other benchmarks. Green values indicate positive residual $r$; red values indicate negative.}
\label{fig:partial}
\end{figure}

Projecting out the first principal component from the standardized 5-benchmark matrix exposes the correlation structure of the residuals (Figure~\ref{fig:partial}). If a single $G$-factor fully explains the positive manifold, the residual correlations should be near-zero and randomly signed. If group factors exist beneath $G$, the residuals will show systematic positive correlations within groups and negative correlations between groups---the signature of a hierarchical factor structure.

It is found that 7 of 10 pairwise residual correlations are \emph{negative}, with a mean residual $r = -0.24$. The positive manifold in the raw correlations (all 15 positive) is largely attributable to $G$. Once $G$ is removed, the residual structure is predominantly \emph{anti-correlated}. This is the pattern for a strong single-factor model with group factors. After removing the shared variance, benchmarks within the same group remain positively correlated, while benchmarks in different groups become negatively correlated revealing suppression in the analysis whereby $G$ masks the group-level structure. Two group factors emerge clearly from the residual matrix:

\begin{enumerate}[label=\textbf{Group \Roman*:}, leftmargin=3em]
    \item {Reasoning:} MATH and GPQA Diamond are strongly positively correlated in the residuals ($r_{\text{resid}} = +0.59$). These benchmarks test multi-step reasoning at the difficulty frontier. MMLU is no longer part of this cluster: its residual correlations with MATH ($-0.35$) and GPQA ($-0.15$) are negative, indicating that MMLU functions as an isolated benchmark once $G$ is removed.
    \item {Execution/Fluency:} GSM8K and HumanEval retain a strong positive residual correlation ($r_{\text{resid}} = +0.53$). These benchmarks are related to procedural execution, step-by-step arithmetic, or code synthesis.
\end{enumerate}

Cross-group correlations are negative: GSM8K$\times$GPQA ($r_{\text{resid}} = -0.80$), HumanEval$\times$GPQA ($r_{\text{resid}} = -0.68$), MMLU$\times$HumanEval ($r_{\text{resid}} = -0.40$). This strong suppressor structure confirms that $G$ is important since it is the primary reason benchmarks appear positively correlated in the raw data.

\subsection{Adjudicating Statistical vs Mechanistic $G$}

The most direct test of whether $G$ is statistical (Outcome~1) or mechanistic (Outcome~2) is to ask whether the group factor structure beneath $G$ is stable across epochs. If the same benchmarks cluster together regardless of which epoch is examined, the structure reflects a genuine computational dissociation. If the clusters rearrange, the structure is an artifact of the particular model population.

A methodological subtlety is important here. Test~(ii) demonstrated that the $G$-factor \emph{rotates} across epochs: the first eigenvector changes direction as new models enter the window. This means that subtracting a global PC1 (computed across all models) from epoch-specific data conflates different $G$-factors---it removes too much variance along directions that are not the epoch's own $G$, and too little along directions that are. The correct procedure is to extract PC1 \emph{within each epoch's data} and subtract only that epoch's $G$ before examining the residual structure.

\begin{figure}[H]
\centering
\includegraphics[width=0.95\textwidth]{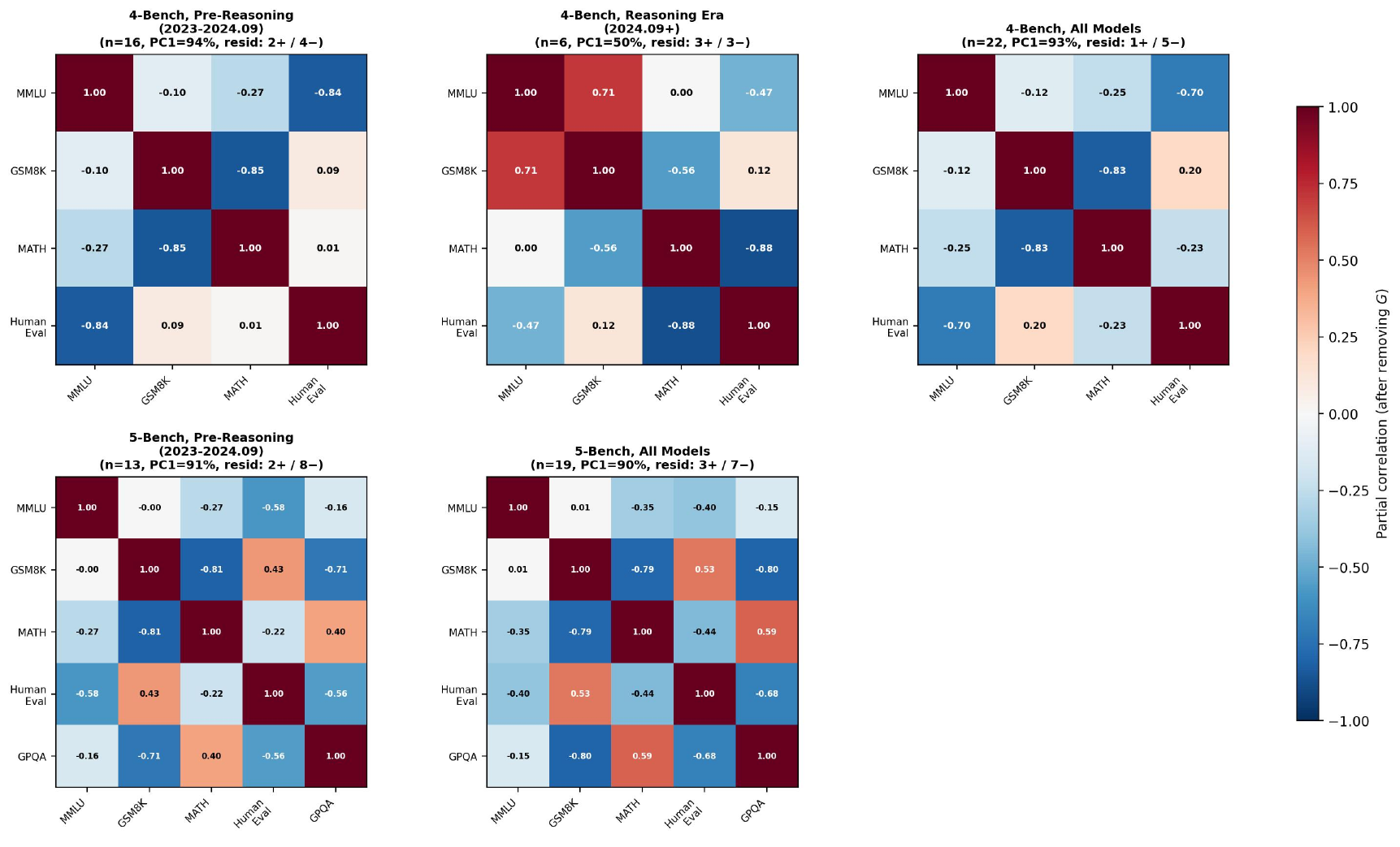}
\caption{\textbf{Epoch-specific partial correlations.} Partial correlation matrices after removing \emph{each epoch's own PC1} from the 4-benchmark battery. (a)~Epoch~II (2023.03--2024.03, $n=8$, $\rho_1 = 92\%$), (b)~Epoch~III (2024.04--2024.09, $n=7$, $\rho_1 = 80\%$). Additional panels show the 5-benchmark and global analyses. Of 6 benchmark pairs in the 4-benchmark analysis, 4 maintain the same sign across Epochs~II and III; only the two weakest pairs flip sign.}
\label{fig:era_comparison}
\end{figure}

\paragraph{4-benchmark analysis.} Partitioning the 4-benchmark battery into the four epochs defined in Section~4.1, Epochs~I ($n=3$) and IV ($n=4$) have insufficient data ($n < K + 1 = 5$), but Epochs~II ($n=8$, $\rho_1 = 92\%$) and III ($n=7$, $\rho_1 = 80\%$) provide well-powered comparisons (Figure~\ref{fig:era_comparison}). Of 6 benchmark pairs, 4 maintain the same sign across Epochs~II and III; only the two weakest pairs (GSM8K$\times$HumanEval and MATH$\times$HumanEval, both $|r_{\text{resid}}| < 0.3$) flip sign. The stable negatives are MMLU$\times$GSM8K, MMLU$\times$HumanEval, MMLU$\times$MATH, GSM8K$\times$MATH which persist across both epochs, indicating that the trade-off structure beneath $G$ is a genuine feature of the model population, not an artifact of any particular temporal window.

\paragraph{5-benchmark analysis.} The expanded GPQA Diamond coverage permits epoch-specific partial correlation analysis on the 5-benchmark battery ($K=5$).  In Epoch~II, the residual correlations are overwhelmingly negative (2 positive, 8 negative). The dominant positive correlation is MATH$\times$GPQA ($r_{\text{resid}} = +0.92$)---competition mathematics and PhD-level science are tightly coupled once $G$ is removed. The second positive correlation is GSM8K$\times$HumanEval ($r_{\text{resid}} = +0.41$), confirming the execution cluster. The cross-group negatives are strong: GSM8K$\times$GPQA ($r_{\text{resid}} = -0.74$), MMLU$\times$HumanEval ($r_{\text{resid}} = -0.64$). The group structure is crisp: a reasoning cluster (MATH, GPQA) and an execution cluster (GSM8K, HumanEval), with MMLU negatively correlated with nearly everything.

In Epoch~III, the pattern shifts. The MATH$\times$GPQA correlation collapses from $+0.92$ to $+0.13$---the tight reasoning cluster has splintered. Meanwhile, the number of positive residual correlations rises from 2 to 4: MATH$\times$HumanEval flips from $-0.24$ to $+0.14$, and MMLU$\times$GPQA flips from $-0.18$ to near zero ($+0.02$). The execution cluster weakens: GSM8K$\times$HumanEval drops from $+0.41$ to $+0.32$. But the cross-group negatives remain stable: GSM8K$\times$GPQA is $-0.77$ (was $-0.74$), and HumanEval$\times$GPQA is $-0.54$ (was $-0.48$).

The interpretation is that the group structure beneath $G$ is partly stable and partly epoch-dependent. The stable features---the strong negative correlation between execution benchmarks (GSM8K) and reasoning benchmarks (GPQA), the negative correlation between MMLU and HumanEval---persist across both epochs and likely reflect genuine computational dissociations in the transformer architecture. The unstable features, captured by the splintering of the MATH$\times$GPQA reasoning cluster reflects the changing model population as labs begin to specialize. In Epoch~II, all models used the same scaling recipe, so MATH and GPQA moved in lockstep; in Epoch~III, mixture-of-experts and early tool augmentation began to decouple these tasks. The verdict is mixed: the broad group structure (reasoning vs.\ execution) is mechanistic, but the fine-grained within-group couplings are population-dependent.

\subsection{Detrending $G$}
\label{sec:detrending}

For each benchmark $b_j$, a linear regression $X_{ij} = \alpha_j + \beta_j \cdot t(m_i) + \varepsilon_{ij}$ is fitted, where $t(m_i)$ is the release date in days. The temporal trend accounts for between 47\% (GSM8K) and 72\% (GPQA Diamond) of variance in individual benchmark scores, with slopes ranging from $+9\%$/year (MMLU) to $+34\%$/year (MATH). The residuals $\hat{\varepsilon}_{ij}$ capture how much each model deviates from the expected score for its release date---whether it is better or worse than the temporal trend predicts.

 On the 5-benchmark battery ($n = 19$): all 10 pairwise correlations among the detrended residuals are positive, with a mean of $\bar{r}_{\text{detrend}} = 0.71$ (compared to $\bar{r}_{\text{raw}} = 0.82$). PC1 still explains 77\% of variance, down from 90\%---a reduction of 13 percentage points. Horn's parallel analysis on the detrended data confirms that PC1 remains significant ($p < .0001$) and PC2 does not ($p = 1.0$).

On the 4-benchmark battery ($n = 22$), the pattern is the same: all 6 pairwise detrended correlations are positive (mean $\bar{r}_{\text{detrend}} = 0.75$), PC1 explains 81\% of variance (down from 93\%), and Horn's analysis confirms significance ($p < .0001$).

The detrended correlations reveal the same group structure as the partial correlations. The strongest detrended pair is GSM8K$\times$HumanEval ($r = +0.90$), confirming the execution cluster. MATH$\times$GPQA ($r = +0.86$) confirms the reasoning cluster. The weakest detrended correlation is GSM8K$\times$GPQA ($r = +0.47$), the cross-group pair. MMLU correlates positively with all benchmarks after detrending ($r = +0.76$ to $+0.79$), suggesting that its isolation in the partial correlation analysis reflects its relationship to $G$ rather than to the residual structure.

The detrended $G$ (77\%) is comparable to the 66\% reported by Ili\'c and Gignac \citep{ilic2024evidence} on a cross-sectional sample of 591 models.

\section{Discussion}

\begin{figure}[H]
\centering
\includegraphics[width=0.8\textwidth]{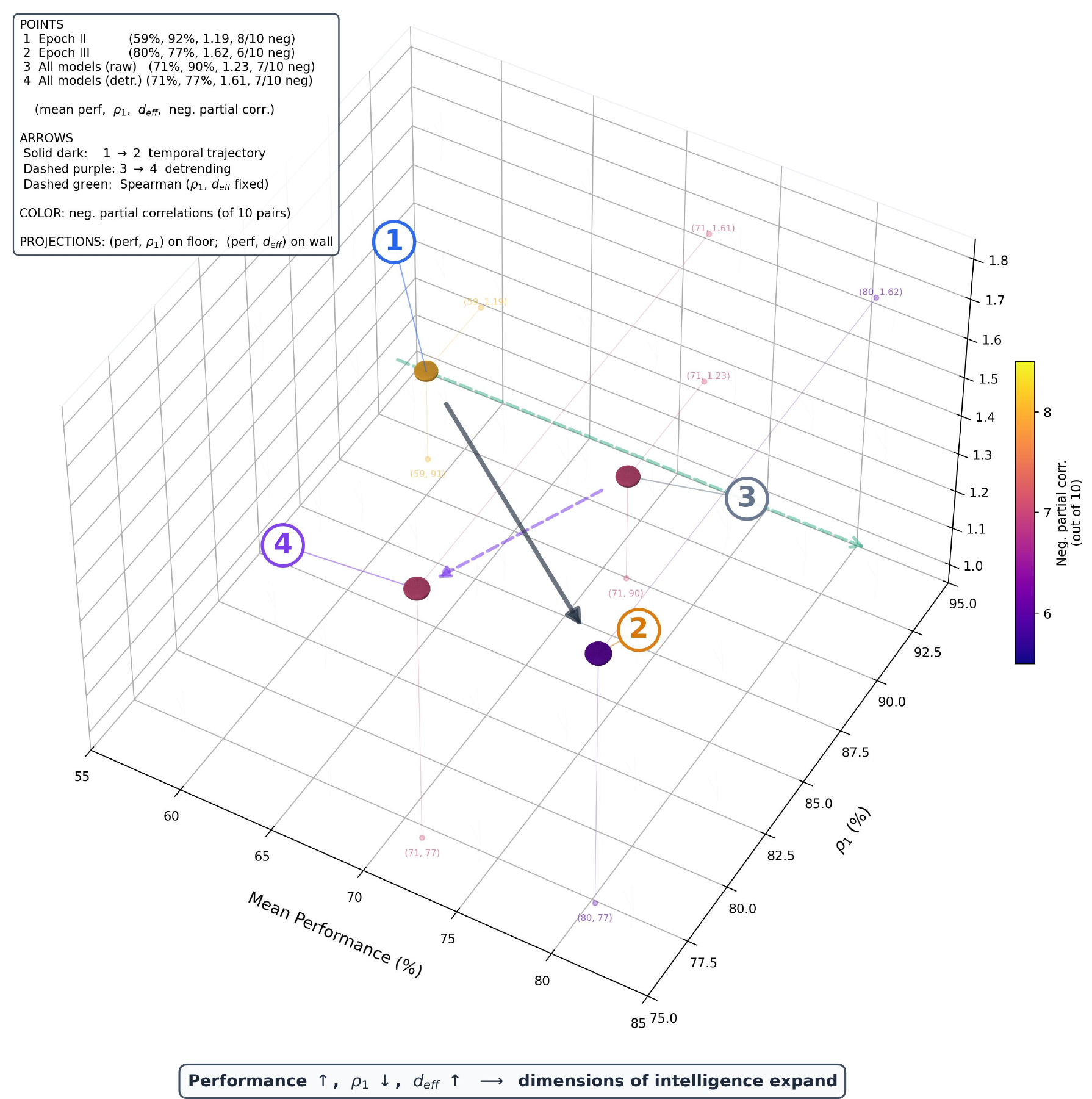}
\caption{\textbf{Trajectory through $G$-space}. Four points are plotted in a space defined by mean benchmark performance, $\rho_1$ (variance explained by $G$), and effective dimensionality $d_{\text{eff}}$. Point~1: Epoch~II (within-epoch, 5-benchmark battery). Point~2: Epoch~III (within-epoch). Point~3: all models, raw PCA. Point~4: all models, detrended PCA. Solid arrow: temporal trajectory (1$\to$2). Dashed purple arrow: detrending correction (3$\to$4). Dashed green line: Spearman prediction (performance increases, $\rho_1$ and $d_{\text{eff}}$ fixed). Sphere colour encodes the number of negative partial correlations out of 10 benchmark pairs. Projection lines show coordinate pairs on the floor (performance, $\rho_1$) and back wall (performance, $d_{\text{eff}}$).}
\label{fig:trajectory}
\end{figure}

Figure~\ref{fig:trajectory} summarizes the principal findings in a three-dimensional space defined by mean benchmark performance, the variance explained by $G$ ($\rho_1$), and the effective dimensionality of the benchmark space ($d_{\text{eff}}$). Four points are plotted: the within-epoch factor structures for Epoch~II and Epoch~III, and the all-models factor structure before and after removing the linear time trend. If the positive manifold behaved as Spearman's $g$ predicts, a fixed latent factor that accounts for correlated improvement, the trajectory would follow the dashed green line where performance increases while $\rho_1$ and $d_{\text{eff}}$ remain constant. Instead, the observed trajectory (solid arrow, point~1 to point~2) moves in the opposite direction, as performance rises from 59\% to 80\%, $\rho_1$ falls from 92\% to 77\%, and $d_{\text{eff}}$ rises from 1.19 to 1.62. The dashed purple arrow shows that detrending the all-models data (point~3 to point~4) reproduces this shift whereby removing the temporal trend moves the all-models point from the Epoch~II structural region ($\rho_1 = 90\%$, $d_{\text{eff}} = 1.23$) to the Epoch~III region ($\rho_1 = 77\%$, $d_{\text{eff}} = 1.61$). Sphere colour encodes the number of negative partial correlations (out of 10 benchmark pairs) after removing each analysis's own $G$, indicating the strength of the suppressor structure beneath the positive manifold. The projection lines onto the floor and back wall display coordinate pairs for each point. The dimensions of capability are expanding and diversifying they are not collapsing onto a single factor. The psychometric dynamics of AI are not toward AGI but something more interesting involving the outsourcing of tools in order to explore a higher dimensional space of capability. LLMs by virtue of the collective archive on which they are trained, and the technological-tool niche in which they live, show evidence of becoming a society of minds. 

\subsection{The Primacy of the Positive Manifold}

The positive manifold is confirmed by all 28 pairwise correlations across 8 benchmarks returning positive values with a mean of $\bar{r} = 0.82$. A single dominant eigenvalue, $G$, captures 90\% of variance in the 5-benchmark core battery ($n = 19$) and 93\% in the 4-benchmark battery. This much is consistent with the psychometric precedent. But what does $G$ actually measure? $G$ loads most heavily on benchmarks that test knowledge-intensive reasoning at the difficulty frontier (MMLU, GPQA, MMLU-Pro, MATH, all at $+0.44$ to $+0.47$), and loads more weakly on procedural execution (HumanEval at $+0.20$). $G$ is not ``being good at everything.'' It is the capacity to generalize across knowledge-informed problems. The residual structure beneath $G$ reveals two group factors and an isolate: a reasoning cluster (MATH, GPQA: $r_{\text{resid}} = +0.59$), an execution cluster (GSM8K, HumanEval: $r_{\text{resid}} = +0.53$), and MMLU, which is largely isolated once $G$ is removed---near-zero residual correlation with GSM8K ($+0.01$) and negative with MATH ($-0.35$) and HumanEval ($-0.40$). Cross-group correlations are strongly negative (GSM8K$\times$GPQA: $-0.80$; HumanEval$\times$GPQA: $-0.68$). The detrending analysis (Section~4.9) confirms that this structure is not a temporal artifact: the positive manifold survives completely after removing the linear time trend, with PC1 still significant at $p < .0001$ and all pairwise correlations remaining positive. The picture has affinities with the Cattell--Horn--Carroll hierarchy in human psychometrics \citep{cattell1963theory,ilic2024evidence}.

\subsection{The Rise and Fall of $G$}

$G$-scores increase monotonically with release date. This is unsurprising and largely uninformative: later models are better at everything because they are engineered to be. The expanding-window analysis confirms that $G$ dominates throughout, with $\rho_1 > 90\%$ across the entire timeline on the 4-benchmark battery. However, the expanding window pools all models cumulatively, so the early models tend to dominate and buffer the late decline revealed by the segmented epoch-specific analysis, which isolates each period and reveals evidence of a decline. 

The detrending analysis (Section~4.9) resolves this apparent tension and clarifies the trajectory. The raw all-models $G$ ($\rho_1 = 90\%$ on the 5-benchmark battery) is a composite of two contributions: genuine shared structure and temporal inflation from all models improving together. Detrending removes the second and leaves $\rho_1 = 77\%$. This figure matches the within-epoch Epoch~III value (also 77\%), and the agreement is not coincidental. Within an epoch, models span only a few months, so the temporal trend is negligible and within-epoch PCA is approximately equivalent to detrended PCA. The correspondence confirms that both methods are measuring the strength of $G$ after the shared trajectory of improvement has been removed.

The trajectory that emerges is:

\begin{center}
\begin{tabular}{lccl}
\toprule
\textbf{Analysis} & $\rho_1$ & $n$ & \textbf{Interpretation} \\
\midrule
Epoch II, within-epoch (5-bench) & 92\% & 7 & Genuine $G$ at its peak \\
Epoch III, within-epoch (5-bench) & 77\% & 7 & Genuine $G$ after specialization begins \\
All models, detrended (5-bench) & 77\% & 19 & Current state of the field, trend removed \\
All models, raw (5-bench) & 90\% & 19 & Genuine $G$ + temporal inflation \\
\bottomrule
\end{tabular}
\end{center}

The within-epoch decline from 92\% to 77\% cannot be attributed to temporal confounding, because there is almost no temporal trend within a six-month epoch. That 15 percentage-point decline is real architectural divergence---models in Epoch~III genuinely share less of their variance than models in Epoch~II. And the fact that the detrended all-models figure matches the Epoch~III within-epoch figure tells us that the current state of the field, after the temporal tide is drained, is already at the Epoch~III level of coherence, not the Epoch~II level. The raw 90\% was flattering because it was inflated by the shared trajectory of improvement.

The rise (to 92\%) occurred during the scaling era, when all labs pursued the same recipe of larger dense transformers trained on more data. The fall (to 77\%) reflects  architectural divergence as labs began to specialize: mixture-of-experts architectures, inference-time reasoning chains, tool augmentation, and code-specialized fine-tuning. The ``rise and fall of $G$'' is therefore better supported than the expanding window alone might have suggested and it is the detrended and within-epoch analyses, not the raw expanding window, that reveal it. 

\subsubsection{The Dawn of a New Dimension}

In Epoch~IV (2024.09+), the second eigenvalue increases ($\lambda_2 = 1.88$), and the dominance ratio $\delta = \lambda_1/\lambda_2$ drops from 15:1 to 1.8:1. The benchmark space is suggestive of departure from 1-dimension. The second component separates ``depth of search'' (MATH, GPQA) from ``breadth of recall'' (MMLU, GSM8K), capturing the distinction between models that invest inference-time compute in reasoning chains (o1, DeepSeek~R1) and those that rely on training-time knowledge. There is at least descriptive evidence that the effective dimensionality rises from $d_{\text{eff}} = 1.3$ to $d_{\text{eff}} = 1.9$ on the 5-benchmark battery.  In \ref{app:parallel} where a full permutation analysis of statistical significance is performed PC1 remains dominant, albeit declining,  but the importance of PC2 cannot be confirmed.  At this point a second dimension remains suggestive and awaits more benchmark data.

\subsection{Tectonic shifts}

The more interesting observation is what happens to the internal structure of $G$ over time.
During Epoch~II (2023.03--2024.03), the component loadings are nearly uniform: all four benchmarks in the core battery load between $+0.48$ and $+0.51$ on PC1, and a single component captures 92\% of variance. The models of this period, including GPT-4, Claude~3, Gemini Ultra, Llama~3, all improve in lockstep across all tasks. $G$ is essentially a scalar and the only thing that varies between models is how far along the common trajectory they have traveled.

By Epoch~III (2024.04--2024.09), the loadings begin to shift. HumanEval's loading on PC1 rises to $+0.54$ above the knowledge benchmarks. The advent of code-specialized fine-tuning and function-calling capabilities begins to decouple procedural execution from broad knowledge. By the time the 6-benchmark battery is examined, HumanEval's loading on $G$ has dropped to $+0.20$, while the knowledge-heavy benchmarks cluster at $+0.44$ to $+0.47$.

\subsubsection{The Great Rotation}

The most consequential result is not a splintering of the eigenvalue spectrum (which is statistically underpowered) but the rotation of the first eigenvector. When DeepSeek~V3 enters the expanding window, the angular displacement of $G$'s eigenvector reaches $6.4°$, which is an order of magnitude larger than any previous step.  What it might mean to be generally intelligent, as operationalized by the benchmark battery, is different after the arrival of tool-augmented and reasoning-chain models than it was before. 

This rotation is the factor-analytic signature of a change in representational basis. In the scaling era, $G$ pointed uniformly across all benchmarks and the general component was to be a bigger transformer trained on more data. In the tools era, $G$ rotates toward knowledge-intensive benchmarks and away from procedural execution because the tools themselves (code interpreters, web search, reasoning scaffolds) handle procedural tasks. This frees the model's own capacity for the problem of generalization across knowledge domains. The reasoning profile of intelligence has changed, and it has recapitulated a very human tendency, to diminish the self through outsourcing. 

\subsection{AI Foxes and Hedgehogs}

The partial correlations, obtained by subtracting each epoch's own PC1, expose what lies beneath the general component.  Four of the six benchmark pairs maintain the same sign across Epochs~II and III, and the two that flip are near-zero in at least one epoch. The stable negative correlations (MMLU$\times$HumanEval, GSM8K$\times$MATH, MMLU$\times$GSM8K, MMLU$\times$MATH) reveal a pattern: once $G$ is removed, increasing capability on one dimension comes at the expense of another. Knowledge trades off against code amd grade-school arithmetic trades off against competition mathematics. The general component masks the underlying specializations that the partial correlations reveal.

This finding has direct implications for the AGI debate. In the scaling era, the appearance of generality was maintained because all capabilities rose together on a single component. In the tools era, the partial correlations show increasing anti-correlation whereby models that excel at reasoning-chain benchmarks (MATH, GPQA) do so partly at the expense of rote procedural execution (GSM8K, HumanEval).  Later models conceal what Isiah Bering in 1951 might have described as skulk of foxes that know many things within a hedgehog that knows one big thing \citep{berlin2013hedgehog}. Unlike Jensen's idea of an underlying `distillate' of intelligence, the models reveal something more akin to Minsky's Society of Mind \citep{minsky1986society}.  

\subsection{Inversion of the Ptolemaic Succession}

There is a useful inverted analogy in the evolution of LLMs to the history of astronomy. Ptolemy's geocentric model could accommodate each new planetary observation by adding another epicycle---another circular motion layered on top of the existing ones. Each addition improved the fit, but the dimensionality of the model grew without bound, and the framework never arrived at a simple underlying law. The current benchmarking regime has a similar structure. Each time a model demonstrates a new capability (tool use, code execution, web browsing, chain-of-thought reasoning), a new benchmark is introduced to measure it. The battery grows, and the component structure becomes more complicated. The positive manifold holds across an expanding set of tasks, with each requiring its own epicycle of evaluation.

The Keplerian move would be to find the \emph{right basis}---a small set of latent dimensions that account for the observed covariance without requiring a new benchmark for each new capability. The Newtonian move would be to move beyond a parsimonious basis by discovering a unifying \emph{law} of transformers that govern how those components evolve across epochs. This is what one might describe as the `Ptolemaic Succession' familiar from the history of natural scientific revolutions.  The inversion of the Ptolemaic Succession in the evolution of LLMS shows how epicycles are not necessarily deleterious given enough computational power. LLMS are able to add epicycle-like specializations (revealed by the partial correlations) and coordinate them through a dominant compressor (the leading positive manifold). If  bottlenecks of the human mind are overcome, as they seem to be in certain domains of LLM application, there is little stopping intelligences becoming more diverse and significantly more complicated, albeit less attractive. 

\subsection{Intelligence is Tool-Using Intelligence}

A final observation concerns the coherence of the benchmarking enterprise itself. The benchmarks in the current battery were designed to test models without tool access as a means of obtaining something close to raw cognitive performance on knowledge retrieval, mathematical reasoning, and code generation. But the models of Epoch~IV do not operate this way. They use tools including code interpreters and search engines. Evaluating a tool-augmented model on a no-tools benchmark is like measuring the intelligence of a literate human by forbidding them to write anything down.

If the benchmarks that define $G$ in AI are to be compared against human reasoning, then human reasoning should also be evaluated with its characteristic tools. Most of what makes \emph{Homo sapiens} cognitively distinctive---language, mathematics, writing, scientific reasoning, institutional knowledge---emerged after the Paleolithic, and therefore after the human brain ceased to evolve in any substantial way \citep{tattersall2012,neubauer2018evolution}. The cognitive explosion of the last 50,000 years is not a story of neural hardware improvement but of tool accumulation: symbolic systems, notational technologies, social institutions, and information storage devices that progressively externalized cognitive functions \citep{clark1998,hutchins2000distributed}. 

The same is now true of LLMs. A model equipped with a code interpreter, a calculator, and a web browser is not the same cognitive system as the same model running in isolation. Its effective intelligence has been extended by its tools, just as human intelligence has been extended by writing, libraries, and the internet. And now these tools include LLMs themselves. In this regime, the concept of a pure general intelligence factor, whether $g$ or $G$, ceases to be well-defined. Intelligence is not a property of the individuated substrate it is a property of the expanded individual-tool system. 

The practical consequence is that benchmarks designed for the pre-tool era of LLMs will progressively lose their meaning as tools become standard. The score matrix spectrum documented in this paper is a snapshot of a transitional period and perhaps the last moment at which it made sense to evaluate models on isolated cognitive tasks. This argues for  less emphasis to be placed on general intelligence, or AGI, which is rooted in psychometric simplifications, and for a concerted effort to reveal the full dimensionality of intelligence. Toward respecting the many different reasoning dimensions of humans, non-human life, and machines \citep{vallor2024ai}. 

\section{Acknowledgements} Thanks to all the participants of the SFI working group on Cognitive Science Perspectives on AGI; and thanks to Melanie Mitchell, John Krakauer, and the Wicklow Foundation for their ongoing leadership and support of  SFI Working Groups on Natural and Artificial Intelligence. D.~C.~K.~ is supported in this work by the Templeton World Charity Foundation, Inc. (funder DOI 501100011730) under the grant DOI:\href{https://doi.org/10.54224/20650}{10.54224/20650} no.\,20650 on “Building Diverse Intelligences through Compositionality and Mechanism Design”,

\clearpage
\appendix

\section{Data}
\label{app:data}

\subsection{Data Structure}

 The score matrix contains $N = 39$ rows (models) and $K = 14$ benchmark columns plus 3 metadata columns (model name, organization, release date), spanning February 2019 to December 2025. An additional column records parameter count where available ($n = 13$ models). Of the $39 \times 14 = 546$ benchmark cells, 246 are populated (42\% overall coverage). The sparsity is structured rather than random: early models (pre-2023) lack scores on benchmarks that did not yet exist (GPQA, MMLU-Pro, MATH-500, IFEval, SWE-bench), while late models (post-2024) often lack scores on older benchmarks (HellaSwag, ARC, WinoGrande) that are no longer routinely reported.

The 39 models represent 7 organizations: OpenAI (10 models: GPT-2 through GPT-5), Anthropic (11 models: Claude~1 through Claude~Opus~4.5), Google (8 models: PaLM through Gemini~3~Pro), Meta (5 models: LLaMA-65B through Llama~4~Maverick), DeepSeek (2: V3, R1), Mistral (2: 7B, 8x7B), and DeepMind (1: Chinchilla). Coverage is highest for MMLU (30/39, 77\%) and lowest for IFEval (4/39, 10\%). The 5-benchmark core battery (MMLU, GSM8K, MATH, HumanEval, GPQA Diamond) yields 19 complete-case models; the 4-benchmark battery (dropping GPQA) yields 22.

\subsection{Sources}

All benchmark scores were compiled from the following source types, in order of priority:

\begin{enumerate}[nosep]
    \item \emph{Official technical reports and model cards.} Primary source for most entries.
    \begin{itemize}[nosep]
        \item \emph{OpenAI:} GPT-2 paper (Radford et al., 2019); GPT-3 paper (Brown et al., 2020); GPT-4 technical report (2023); GPT-4o, GPT-4.5, and GPT-5 system cards (2024--2025); o1 system card (2024); o3 technical report (2025).
        \item \emph{Anthropic:} Claude~1 and Claude~2 model cards (2023); Claude~3 model card covering Opus, Sonnet, and Haiku (2024); Claude~3.5~Sonnet model cards, June and October releases (2024); Claude~3.7~Sonnet, Claude~Opus~4, Claude~Sonnet~4, and Claude~Opus~4.5 model cards (2025).
        \item \emph{Google/DeepMind:} PaLM technical report (Chowdhery et al., 2022); Chinchilla paper (Hoffmann et al., 2022); PaLM~2 technical report (2023); Gemini~1.0 technical report (2023); Gemini~1.5 technical report (2024); Gemini~2.0~Flash and Gemini~2.5~Pro blog posts (2024--2025); Gemini~3~Pro technical report (2025).
        \item \emph{Meta:} LLaMA paper (Touvron et al., 2023); Llama~2 paper (Touvron et al., 2023); Llama~3 paper (2024); Llama~3.1~405B technical report (2024); Llama~4~Maverick technical report (2025).
        \item \emph{DeepSeek:} DeepSeek-V3 technical report (2024); DeepSeek-R1 technical report (2025).
        \item \emph{Mistral:} Mistral~7B paper (Jiang et al., 2023); Mixtral~8x7B technical report (Jiang et al., 2024).
    \end{itemize}
    \item \emph{Published benchmark papers.} Used for verification and for scores not reported in model cards. Examples: Hendrycks et al.\ (2021) for MMLU and MATH baselines, Chen et al.\ (2021) for HumanEval, Rein et al.\ (2024) for GPQA.
    \item \emph{Third-party evaluation platforms.} Epoch AI's AI Benchmarking Hub \citep{epochai2024} was used as a cross-reference for scores reported across multiple sources, and was the primary source for GPQA Diamond scores on models whose own technical reports did not include this benchmark. The LMSYS Chatbot Arena and the Open LLM Leaderboard were consulted but not used as primary score sources, as their evaluation protocols differ from the standard benchmark implementations.
\end{enumerate}

\section{Horn's Analysis}
\label{app:parallel}

A standard objection to PCA-based factor retention at small sample sizes is that the Kaiser criterion ($\lambda > 1$) overcounts factors. Horn's parallel analysis \citep{horn1965} provides a more rigorous alternative. Observed eigenvalues are compared against those obtained from $B = 10{,}000$ random permutations of the data matrix (each column permuted independently, destroying any correlation structure while preserving marginal distributions). A factor is retained only if its observed eigenvalue exceeds the 95th percentile of the corresponding permuted eigenvalue distribution.

\begin{table}[H]
\centering
\caption{\textbf{Permutation Analysis} For each analysis, the observed first and second eigenvalues are compared against the 95th and 99th percentiles of the permutation null distribution ($B = 10{,}000$). A factor is retained if $\lambda_{\text{obs}} > \lambda_{95}$. Exact $p$-values are the fraction of permuted eigenvalues exceeding the observed value.}
\label{tab:parallel}
\begin{tabular}{lcccccc}
\toprule
\textbf{Analysis} & $n$ & $K$ & $\lambda_1^{\text{obs}}$ & $\lambda_1^{95\text{th}}$ & $p(\lambda_1)$ & \textbf{PC1 retained?} \\
\midrule
4-bench, all models & 22 & 4 & 3.73 & 1.81 & $<.0001$ & Yes ($p < .01$) \\
5-bench, all models & 19 & 5 & 4.50 & 2.06 & $<.0001$ & Yes ($p < .01$) \\
6-bench, all models & 9 & 6 & 3.73 & 2.91 & $.0002$ & Yes ($p < .01$) \\
Epoch II, 4-bench & 8 & 4 & 3.66 & 2.44 & $<.0001$ & Yes ($p < .01$) \\
Epoch II, 5-bench & 7 & 5 & 4.57 & 2.89 & $<.0001$ & Yes ($p < .01$) \\
Epoch III, 4-bench & 7 & 4 & 3.21 & 2.56 & $.0005$ & Yes ($p < .01$) \\
Epoch III, 5-bench & 7 & 5 & 3.87 & 2.91 & $.0001$ & Yes ($p < .01$) \\
\bottomrule
\end{tabular}

\vspace{4pt}
\begin{tabular}{lcccccc}
\toprule
\textbf{Analysis} & $n$ & $K$ & $\lambda_2^{\text{obs}}$ & $\lambda_2^{95\text{th}}$ & $p(\lambda_2)$ & \textbf{PC2 retained?} \\
\midrule
4-bench, all models & 22 & 4 & 0.16 & 1.30 & $1.0$ & No \\
5-bench, all models & 19 & 5 & 0.35 & 1.48 & $1.0$ & No \\
6-bench, all models & 9 & 6 & 1.49 & 1.92 & $.65$ & No \\
Epoch II, 4-bench & 8 & 4 & 0.24 & 1.48 & $1.0$ & No \\
Epoch II, 5-bench & 7 & 5 & 0.22 & 1.79 & $1.0$ & No \\
Epoch III, 4-bench & 7 & 4 & 0.42 & 1.51 & $1.0$ & No \\
Epoch III, 5-bench & 7 & 5 & 0.56 & 1.80 & $1.0$ & No \\
\bottomrule
\end{tabular}
\end{table}

\begin{enumerate}

\item The observed first eigenvalue exceeds the 99th percentile of the permutation null in all seven analyses. Even in the most challenging case---the 6-benchmark battery with only $n = 9$ models---the general factor is significant at $p = .0002$. 

\item No second eigenvalue approaches the 95th percentile of its permutation null in any analysis. The Kaiser criterion ($\lambda > 1$) would have retained a second factor in the 6-benchmark battery ($\lambda_2 = 1.49$), but parallel analysis does not ($\lambda_{95} = 1.92$). 

\item  Both Epoch~II and Epoch~III show a significant single factor on both the 4-benchmark and 5-benchmark batteries. The 5-benchmark epoch-specific results are new: the expanded GPQA Diamond coverage (from Epoch~AI's independent evaluations) now permits within-epoch parallel analysis on the richer battery. Epoch~II yields $\rho_1 = 92\%$ on the 5-benchmark battery ($p < .0001$) and Epoch~III yields $\rho_1 = 77\%$ ($p = .0001$). Since models within an epoch span only a few months, the temporal confound is minimal. The positive manifold holds among near-contemporaneous models on the full 5-benchmark battery---the strongest evidence against a purely statistical $G$.

\item  With $n = 4$ models and $K = 4$ benchmarks, the Epoch~IV correlation matrix has zero degrees of freedom and parallel analysis cannot be applied. The two-factor structure reported for Epoch~IV (Section~4.2) remains a descriptive observation.

\end{enumerate}
\pagebreak

\bibliographystyle{apalike}
\bibliography{refs_gfactor}

\end{document}